\documentclass[a4paper,11pt]{article}
\usepackage{latexsym,amssymb,amsmath}
 \usepackage[english]{babel}

\usepackage{amssymb,dsfont,amsmath,amsfonts,hyperref,mdframed,mathrsfs,stmaryrd}
\usepackage[active]{srcltx}
\usepackage{slashed}
\usepackage{color}
\usepackage{amsthm}
\usepackage{scrtime}
\usepackage{cite}
\usepackage[normalem]{ulem}
\usepackage{framed}
\usepackage{cancel}
\usepackage{multirow}
\usepackage{graphicx}
\usepackage{arydshln}
\usepackage{bbm}

\usepackage{booktabs}

\makeatletter

\@addtoreset{equation}{section}
\makeatother

\usepackage{amsmath}
\usepackage{amsfonts}
\usepackage{units}
\usepackage{color}
\usepackage{slashed}
\usepackage{parskip}

\definecolor{hs}{rgb}{0.0,0.67, 0.2}

\def\be{\begin{equation}}
\def\ee{\end{equation}}
\def\ba{\begin{array}}
\def\ea{\end{array}}
\def\bea{\begin{eqnarray}}
\def\eea{\end{eqnarray}}
\def\nn{\nonumber}

\newcommand{\eq}[1]{(\ref{#1})}

\newcommand{\beqs}{\begin{eqnarray}}
\newcommand{\eeqs}{\end{eqnarray}}

\def\({\left(}
\def\){\right)}

\def\L{ {\cal L}}

\def\mxth{\mathsurround=0pt }
\def\xversim#1#2{\lower2.pt\vbox{\baselineskip0pt \lineskip-.5pt
x  \ialign{$\mxth#1\hfil##\hfil$\crcr#2\crcr\sim\crcr}}}

\renewcommand{\L}{\Lambda}

\def\bfone{\relax{\rm 1\kern-.35em 1}}

\newcommand{\bse}{\begin{subequations}}
\newcommand{\ese}{\end{subequations}}

\definecolor{darkred}{rgb}{0.65,0.15,0}
\hypersetup{pdfborder={0 0 0},colorlinks=true,urlcolor=blue,citecolor=blue,linkcolor=darkred,linktocpage=true}

\def\tv{\widetilde{\varphi}}

\begin{document}

\thispagestyle{empty}

\begin{flushright}\small
MI-HET-854    \\

\end{flushright}


\bigskip
\bigskip

\vskip 10mm

\begin{center}

{\LARGE{\bf Scale separation on AdS$_3\times S^3$ \\[1.2ex]
with and without supersymmetry}}

\vskip 4mm

\end{center}


\def\hA{{\widehat A}}
\def\hB{{\widehat B}}
\def\L+{{\widehat\lambda_{+}}}
\def\P-{{\widehat\psi_{-}}}
\def\etam{\overline{\eta_-}}
\def\ep{\overline{\eta_+}}
\def\cm{\widehat{\chi}_{2-}}
\def\hA{\widehat{A}}
\def\hB{\widehat{B}}
\def\ve{{\varepsilon}}

\vskip 2mm

\begin{center}

{\large Aymeric Proust$^{\dagger,}$\footnote{\tt aymeric.proust@ens-lyon.fr}, Henning Samtleben$^{\dagger,\diamondsuit,}$\footnote{\tt henning.samtleben@ens-lyon.fr}, and Ergin Sezgin$^{\star,}$\footnote{\tt sezgin@tamu.edu}}\\
\vskip 8mm

$^\dagger$\, {\it ENSL, CNRS, Laboratoire de physique, F-69342 Lyon, France} \\

\vskip 3mm
 
 $^\diamondsuit$\ {\it Institut Universitaire de France (IUF)}\\
\vskip 3mm

$^\star$\,{\it George P. and Cynthia W. Mitchell Institute \\for Fundamental
Physics and Astronomy \\
Texas A\&M University, College Station, TX 77843-4242, USA}\\
\vskip 3mm

\end{center}

\vskip1.5cm

\begin{center} {\bf Abstract } \end{center}

Six-dimensional chiral gauged Einstein-Maxwell supergravity admits a two-parameter rotating dyonic string solution whose near horizon limit is the direct product of AdS$_3$ and a squashed three-sphere $S^3$. For a particular relation between the two parameters, the solution preserves $1/2$ supersymmetry.  We determine the complete Kaluza-Klein spectrum of the theory around these AdS$_3$ backgrounds. For the supersymmetric backgrounds, the states organize into infinite towers of long and short multiplets of ${\rm OSp}(2|2)$. 
In a certain limit of parameters, both the supersymmetric and the non-supersymmetric spectra exhibit scale separation. In the latter case there are five topologically massive vectors and five scalars retaining finite masses with integer conformal dimensions, and in the supersymmetric case there are supersymmetric partners with half integer conformal dimensions, while all other masses diverge.

\setcounter{footnote}{0}

\newpage

\tableofcontents

\section{Introduction}

A well known property of AdS compactifications of supergravity theories, regardless of whether they are embedded in string/M theory or not, is that in the absence of orientifolds they tend to give a mass scale of the Kaluza-Klein towers which is at the same order as the AdS cosmological constant.  This lack of separation of scale,  which poses a problem in realistic uses of AdS compactifications, means that one cannot separate the lowest energy modes from the KK modes, and therefore the theory is the original higher dimensional theory in disguise. The often studied scale separation is referred to as the parametric one in the sense that one looks for an infinite family of vacua which shows scale separation as one moves along the family \cite{Tsimpis:2012tu,Gautason:2015tig,Font:2019uva,Lust:2020npd}. For a recent review, see \cite{Coudarchet:2023mfs}. 
While inclusion of orientifolds can give rise to scale separation \cite{Gautason:2015tig,Font:2019uva,Farakos:2020phe,Apers:2022zjx,Farakos:2023wps,Arboleya:2024vnp,VanHemelryck:2025qok,Farakos:2025bwf}, the assumptions made in achieving this have been a subject of debate reviewed in \cite{Coudarchet:2023mfs}.
Scale separation in supersymmetric AdS vacua has also been conjectured to be ruled out \cite{Lust:2019zwm} in the context of Swampland Program where a number of criteria are conjectured for the existence of a consistent theory of quantum gravity, amenable to UV completion, at least in principle. 
The most direct way of studying the scale separation problem is to determine the full Kaluza-Klein spectrum the higher dimensional theory in AdS compactifications. Except for the most symmetric cases, however, this can be a very complicated task. So far, many of the scale separation studies consider supergravities in 10D and 11D associated with string/M theory low energy limits, and the full fledged spectrum computation is bypassed by examining the behavior of entities such as the ratio of the internal and external curvatures, which may involve numerical analysis in certain approaches \cite{Tsimpis:2012tu}. 

In this paper, we take the direct approach of computing the full spectrum of six-dimensional chiral gauged Maxwell-Einstein supergravity, often referred to as the Salam-Sezgin model \cite{Salam:1984cj}, compactified on AdS$_3\times ({\rm squashed}\, S^3)$. 
The gauged symmetry is the ${\rm U}(1)_{RS}$ subgroup of the R-symmetry group ${\rm Sp}(1)_{RS}$. The bosonic part of the Lagrangian is displayed in \eq{L6D}. This model should be viewed as a subsector of a model with extra matter multiplet couplings such that the local gauge, gravitational mixed anomalies are canceled by Green-Schwarz-Sagnotti mechanism. Such anomaly free models are very rare unless the gauge group consists of ${\rm SU}(2)$ and ${\rm U}(1)$ factors alone \cite{Randjbar-Daemi:1985tdc,Avramis:2005qt,Avramis:2005hc,Becker:2023zyb}. Furthermore, their embedding to string/M theory is not known, as further discussed in the conclusions.

 As such it provides an interesting framework for studying the scale separation problem, without having to be tied up with structures inherited from the reductions of supergravities in 10D or 11D. This is in the spirit of the Swampland Program.

Our main findings are as follows. The solution we consider, namely AdS$_3 \times S^3_\xi$ where $\mbox{$\xi=\cosh^2\!\beta$}$ is the squashing parameter, arises as the near horizon geometry of a two-parameter rotating dyonic solution \cite{Ma:2023rdq}, and it takes the form 
\bea
ds^2 &=&  \left(L^2\, ds^2_{\rm{AdS}_3} +\frac14 \,\sigma_3^2 +\frac14 \cosh^2\!\beta \left(\sigma_1^2+\sigma_2^2\right) \right)\ ,
\label{solab3-intro}\\[.5ex]
F&=& \frac1{2g}\,\tanh\alpha\,\tanh\beta\: \sigma_1\wedge \sigma_2\ ,
\nonumber\\[.5ex]
H &=& \frac1{g^2}\,\frac{\tanh^2\!\alpha}{\cosh^2\!\beta} \left( L^2\, \omega_{\rm{AdS}_3} +\omega_{S^3}\right)\ ,\qquad
e^\phi=4\,g^4\,{\coth^4\!\alpha}\,{\cosh^4\!\beta} \ ,
\nonumber
\eea
in terms of the left-invariant 1-forms $\sigma_i$ on the three-sphere,
with constant parameters $(\alpha,\beta)$, and $L=\cosh\alpha\,\cosh\beta$\,.
For generic values of $(\alpha,\beta)$, the solution has no supersymmetry, for nonvanishing $\alpha=\beta \ne 0$ it preserves 1/2 supersymmetry and reproduces the solution found in \cite{Gueven:2003uw}. The metric $ds^2_{\rm{AdS}_3}$ is that of AdS$_3$ with unit radius, and $\omega_{\rm{AdS}_3}$ is the associated volume form.
For $\alpha=\beta=0$, the solution preserves full supersymmetry. The last case requires vanishing gauge coupling $g$ in six dimensions and corresponds to the standard AdS$_3\times S^3$ solution of ungauged six-dimensional supergravity.
Using exceptional field theory techniques we calculate the complete spectrum in this background. The procedure will be briefly recalled below, and the results are summarized and discussed in  section~\ref{sec:spectrum}. The bulk of the paper is devoted to the computation of the complete Kaluza-Klein spectrum. 
In particular, we observe that all scalar modes are perturbatively stable (in the sense that they obey the Breitenlohner-Freedman bound) even around the generic non-supersymmetric background.
Furthermore, it turns out that in a certain limit of parameters, in particular entailing $\alpha \to \infty$, both the supersymmetric and the non-supersymmetric spectra exhibit scale separation, with only ten fields retaining finite masses while all other masses diverge. The presence of the $R$-symmetry gauging is essential for this to happen.
To the best of our knowledge, this seems to be the first example of a scale separation phenomenon in AdS compactifications, in the absence of orientifolds. Moreover, we find that all finite masses correspond to integer conformal dimensions, as observed in other scenarios of scale separation~\cite{Conlon_2022,Apers:2022zjx,Apers_2022,Arboleya:2024vnp,Farakos:2025bwf}.

We shall comment further on these results in the final sections. For recent results on scale separation in AdS$_3$ compactifications with orientifolds involved, see \cite{Farakos:2020phe,Apers:2022zjx,Farakos:2023wps,Arboleya:2024vnp,VanHemelryck:2025qok,Farakos:2025bwf}.
For the supersymmetric family of backgrounds, we find that the full spectrum organizes into infinite towers of both, long and short supermultiplets of the background isometry supergroup ${\rm OSp}(2|2)$, summarized in equations (\ref{eq:AllSpectrumLong})--(\ref{eq:spectrumNvectors}) below. Knowledge of this exact spectrum, especially the protected short multiplets, may also shed light on the still elusive holographic CFT dual of the Salam-Sezgin model.

To compute the full Kaluza-Klein spectrum of the model around the AdS$_3 \times S_\xi^3$ background, we make use of the techniques introduced from exceptional field theory \cite{Malek:2019eaz,Malek:2020yue,Eloy:2020uix}.
As a first step this requires the reformulation of the model (\ref{L6D}) as an exceptional field theory (ExFT). For vanishing six-dimensional gauge coupling $g=0$, this exceptional field theory has been constructed in \cite{Hohm:2017wtr}. Within the bosonic sector, the inclusion of the $R$-symmetry gauging simply amounts to the addition of the corresponding potential term to the ExFT Lagrangian. The ExFT formulation provides an explicit description of the consistent truncation of the six-dimensional theory down to a three-dimensional supergravity. Again, the presence of the potential term does not obstruct the consistency of the truncation thanks to the particular dilaton power in this term, as already observed in \cite{Cvetic:2000dm}. In particular, the uplift formulas from \cite{Eloy:2021fhc} do not depend on the gauge coupling constant $g$.
The consistent truncation together with the ExFT formulation allows to compute the mass spectrum of fields not only for the fields of the consistent truncation but for the full infinite Kaluza-Klein towers. Schematically, the full set of Kaluza-Klein fluctuations can be parametrized as $\Phi^{{\cal A}\Lambda}$, with the index ${\cal A}$ referring to the fields of the three-dimensional supergravity, and the index $\Lambda$ labeling the scalar harmonics on the internal $S^3$. The general mass matrix for the scalar fields then turns out to be of the schematic form
\begin{equation}
    {\cal M}_{{\cal A}\Lambda,{\cal B}\Sigma} = 
    {\cal M}_{{\cal AB}}\,\delta_{\Lambda\Sigma}
    +{\cal N}_{{\cal AB}}{}^{PQ} \,{\cal T}_{PQ,\Lambda\Sigma}
    + \delta_{{\cal AB}}\,{\cal M}_{\Lambda\Sigma}
    \,,
    \label{MM-intro}
\end{equation}
where ${\cal M}_{{\cal AB}}$ is the mass matrix of the three-dimensional theory around AdS$_3$,  ${\cal M}_{\Lambda\Sigma}$ is the mass operator for the spin-2 fluctuations, and the middle term combines an action on the scalar harmonics with an action on the three-dimensional indices~\cite{Malek:2019eaz,Malek:2020yue,Duboeuf:2023cth}. For the vector fields (which in three dimensions satisfy first-order equations) and the fermions, the structure of the mass matrices is of similar form, without the last term of (\ref{MM-intro}) \cite{Eloy:2020uix,Cesaro:2020soq,Eloy:2021fhc}.

The rest of this paper is organized as follows. In section~\ref{sec:6D}, we spell out the six-dimensional model and the two-parameter class of AdS$_3 \times S_\xi^3$ solutions. We describe the consistent truncation to three-dimensions and the resulting ${\cal N}=4$ supergravity in  section~\ref{sec:3D}. In particular, we recover the two-parameter family of solutions as a family of stationary points of the three-dimensional scalar potential.  In section~\ref{sec:N4}, we review the known supersymmetric Kaluza-Klein spectrum for the AdS$_3 \times S^3$ solution of ungauged six-dimensional supergravity, which is recovered as the $\alpha=\beta=0$ limit of the two-parameter background. In section \ref{sec:SpectrumKaluza-Klein} we compute the full spectrum Kaluza-Klein around the general two-parameter background, first within three-dimensional supergravity, and subsequently for the higher Kaluza-Klein modes for all different spins. We summarize the result in section \ref{sec:spectrum}, where we also organize the spectrum for the supersymmetric background into multiplets of the background isometry supergroup ${\rm OSp}(2|2)$. Finally, we close in section~\ref{sec:conclusions} with further comments and an outlook.

\section{The 6D model and the dyonic string solution}
\label{sec:6D}

The bosonic Lagrangian for the Salam-Sezgin model is \cite{Salam:1984cj}
\bea 
\mathcal{L}=R * 1\!\!1-\frac{1}{4} * d \phi \wedge d \phi-\frac{1}{2} e^\phi * H \wedge H -\frac{1}{2} e^{\frac{1}{2} \phi} * F \wedge F -8 g^2 e^{-\frac{1}{2} \phi} * 1\!\!1\ ,
\label{L6D}
\eea
in six dimensions, where $F=dA$ and $H=dB+\frac12 F\wedge A$, and $A$ and $B$ are the $U(1)_R$ gauge field that gauges the R-symmetry, $B$ is the two-form potential, and $\phi$ is the dilaton. 
Note, that the equations of motion are invariant under the rescaling
\begin{equation}
    e^\phi\rightarrow \lambda^{-4}\,e^\phi\;,\quad 
    F\rightarrow \lambda\mu\,F\;,\quad
    H\rightarrow\,\lambda^2\mu^2\,H\;,\quad 
    g_{\mu\nu}\rightarrow \mu^2\,g_{\mu\nu}\;,\quad
    g\rightarrow (\lambda\mu)^{-1}\,g
    \;,
    \label{scaling}
\end{equation}
where $\lambda$ is the $\mathbb{R}^+$ scaling symmetry of the Lagrangian (\ref{L6D}) and $\mu$ is the trombone symmetry.
In particular, the rescaling can be used to put the coupling constant $g$ to an arbitrary value as long as it is non-vanishing. At fixed $g$, the theory still admits a scaling symmetry (\ref{scaling}) with $\lambda=\mu^{-1}$ which only affects the metric and the dilaton.

In this paper, we will study the following two-parameter background solution of (\ref{L6D})
\bea
ds^2 &=&  \left(L^2\, ds^2_{\rm{AdS}_3} +\frac14 \,\sigma_3^2 +\frac14 \,\xi\left(\sigma_1^2+\sigma_2^2\right) \right)\ ,
\label{sol}\\
F&=& k\, \sigma_1\wedge \sigma_2\ ,
\label{sol2}\\
H &=& 2\,q \left( L^2\, \omega_{\rm{AdS}_3} +\omega_{S^3}\right)\ ,\qquad
e^\phi=\frac{1}{q^2}\ ,
\label{sol3}
\eea
with free parameters $k, \xi>0$, and 
\begin{equation}
   q =  \frac{2 \,{k}^2}{\xi-1}, \qquad L^2=\frac{\xi\,(\xi-1)}{\xi\,(1-4\,k^2g^2)-1}
   =\frac{\xi}{1-2\,g^2q\,\xi}
   \ .
   \label{qell}
\end{equation}
The $\sigma_i$ are the left-invariant 1-forms 
on the three-sphere
\be
\sigma_1+i\sigma_2 =e^{-i\psi} \big( d\theta+i\sin \theta d\varphi \big)\ ,\qquad \sigma_3=d\psi+\cos\theta d\varphi\ ,
\ee
in terms of which
\begin{equation}
    ds^2_{S^3} = \frac14\,\big(\sigma_1^2+\sigma_2^2+\sigma_3^2\big)\,,\;\;\mbox{and}
    \quad
    \omega_{S^3} = \frac18\,\sigma_1\wedge\sigma_2\wedge\sigma_3
    \,,
\end{equation}
denote the round $S^3$ of radius 1, and its associated volume form.
The metric
\be
ds^2_{\rm{AdS}_3} = \frac{d\rho^2}{\rho^2} 
+ dx^2-2 \rho^2 dx dt\ ,
\ee
is the metric of AdS$_3$ with unit radius, and $\omega_{\rm{AdS}_3}$ is the associated volume form.
Thus, the metric \eq{sol} is a direct product of $\rm{AdS}_3$ with radius-square $L^2$ and a squashed three-sphere with squashing parameter $\xi$.
Let us note that (\ref{qell}) implies the bound
\begin{equation}
    k^2 \,\le\, \frac{\xi-1}{4\,\xi\,g^2}
    \,,
    \label{eq:bound}
\end{equation}
for the magnetic monopole charge, which furthermore implies the bound $\xi\ge1$ for the squashing parameter.

The solution (\ref{sol})--(\ref{sol3}) arises as the near-horizon of an extremal rotation dyonic string solution \cite{Ma:2023rdq}. For generic values of $\xi$ and $k$, the solution is non-supersymmetric. For special values of the parameters, half of the supersymmetries can be preserved, provided that \cite{Ma:2023rdq}
\begin{equation}
  1/2\  {\rm supersymmetry} \;\;\Longrightarrow\;\; L=\xi\,,
    \label{susy}
\end{equation}
and in this case the solution reduces to the one-parameter solution found in \cite{Gueven:2003uw}. In particular, for the supersymmetric solution $H$ becomes self-dual. To see this, we note that
\be
\ast_6 H = 2\,q \left(\frac{L^3}{\xi} \, \omega_{\rm{AdS}_3} + \frac{\xi}{L}\, \omega_{S^3} \right)\ .
\label{h6}
\ee

For vanishing gauge coupling $g=0$, the solution gains an additional scaling symmetry from (\ref{scaling}), which can be used to set $k=1$, and reproduces the one-parameter solution found in \cite{Eloy:2021fhc}. In this case, (\ref{qell}) yields $L^2=\xi$, i.e.\ according to (\ref{susy}) the solution is not supersymmetric.
Interestingly, in \cite{Eloy:2021fhc} the  background (\ref{sol})--(\ref{sol3}) at $g=0$ was found as a supersymmetric solution, however, within the non-chiral six-dimensional ${\cal N}=(1,1)$ supergravity. In contrast, within the ${\cal N}=(1,0)$ model (\ref{L6D}), for $g=0$, the same solution, for $\xi>1$, breaks all supersymmetries.
This is an illustration of the fact the bosonic model (\ref{L6D}), for $g=0$, and upon coupling three additional vector multiplets, admits two inequivalent fermionic extensions, one with ${\cal N}=(1,0)$, and one with ${\cal N}=(1,1)$ supersymmetry. Their relation can be understood, as both models are obtained as inequivalent truncations of the ${\cal N}=(2,1)$ parent theory \cite{Roest:2009sn}.

Within the ${\cal N}=(1,0)$ model, supersymmetry at $g=0$ requires $L=\xi=L^2$, thus $\xi=1$, i.e.\ absence of squashing. The background in this case reduces to the standard AdS$_3\times S^3$ solution, at which supersymmetry is fully restored
\begin{equation}
\mbox{full supersymmetry} \;\;\Longrightarrow\;\; g=0=\xi-1\,.
    \label{susy-enhancement}
\end{equation}

Finally, let us discuss the limit $\xi\rightarrow1$ for $g\ne 0$, in which the squashing disappears while the internal manifold is given by the round $S^3$.  In this limit, equations (\ref{qell}) imply that $k\rightarrow0$ while $q$ can be kept as a free parameter, related to the AdS$_3$ radius as
\begin{equation}
    L^2=\frac{1}{1-2\,q\,g^2}\,.
\end{equation}
In this case supersymmetry is fully broken.

\section{The 3D model and the vacuum solution}
\label{sec:3D}

The six-dimensional theory (\ref{L6D}) admits a consistent truncation on $S^3$ \cite{Cvetic:2000dm,Deger:2014ofa,Baguet:2015iou,Eloy:2021fhc} 
leading to an ${\cal N}=4$ 
three-dimensional gauged supergravity\footnote{In three dimensions, ${\cal N}=4$ corresponds to 8 real supercharges, descending from ${\cal N}=(1,0)$ supersymmetry in six dimensions.}
whose bosonic sector is described by the Lagrangian
\begin{equation}	 
	e^{-1}\mathcal{L}=R+\frac1{8}g^{\mu\nu}D_\mu M^{PQ}D_\nu M_{PQ}+e^{-1}\mathcal{L}_{\text{CS}}-V\,.
 \label{3DL}
\end{equation}
For $g=0$, i.e.\ no $R$-symmetry gauging in six dimensions, the truncation has been worked out in \cite{Eloy:2021fhc}, after embedding (\ref{L6D}) into the half-maximal ${\cal N}=(1,1)$ supergravity in six dimensions. The additional $g^2$ term in (\ref{L6D}) is consistent with the consistent truncation and simply gives rise to an additional contribution in the scalar potential $V$ \cite{Cvetic:2000dm}, see (\ref{eq:potential}) below. Let us recall that in \cite{Eloy:2021fhc} the model (\ref{3DL}) is found within a half-maximal ${\cal N}=8$ supergravity. Although its bosonic sector is identical to (\ref{3DL}), the fermionic field content and the fermionic couplings differ, in complete analogy to the discussion after (\ref{h6}) for the corresponding six-dimensional models. The fermionic sector for the present model can be worked out by embedding the bosonic Lagrangian (\ref{3DL}) into the general structure of 3D, ${\cal N}=4$ supergravities \cite{deWit:2003ja}.

The scalar target space in (\ref{3DL}) is given by the coset space
\begin{equation}	\label{eq:coset}
	\text{G}/\text{H}=\text{SO}(4,5)/\big(\text{SO}(4)_{\rm o}\times\text{SO}(5)\big)\,,
\end{equation}
parametrized by the symmetric matrix $M^{PQ}$, $P, Q = 1, \dots 9$, whose indices are raised and lowered with the constant $\text{SO}(4,5)$ invariant tensor $\eta_{PQ}$.
The model (\ref{3DL}) has local gauge symmetry
\begin{equation}	\label{eq:gauge}
	\text{G}_{\text{gauge}}=\left[\text{SO}(4)_{\rm gauge}\ltimes(\mathbb{R}^3\times \mathbb{R}^3)\right] \times \mathbb{R}^1 \times {\rm SO}(2)_{RS} \,,
\end{equation}
where $\text{SO}(4)_{\rm gauge}$ corresponds to the isometries of the round $S^3$, the $(\mathbb{R}^3\times \mathbb{R}^3)\times \mathbb{R}^1$ correspond to nilpotent shifts, broken at the vacuum, and the ${\rm SO}(2)_{RS}$ is the gauged $R$-symmetry from six dimensions.
The local symmetry (\ref{eq:gauge}) is implemented by covariant derivatives  
\begin{equation}
	D_\mu M_{PQ}=\partial_\mu M_{PQ}+4\,A_\mu{}^{RS}\,\Theta_{RS\vert(P}{}^{T}\, M_{Q)T}
    +4\,A_\mu{}^{a}\,\Theta_{a\vert(P}{}^{T}\, M_{Q)T}\,,
    \label{covD}
\end{equation}
in terms of vector fields $A_\mu{}^{PQ}=-A_\mu{}^{QP}$, and $A_\mu^a$\,, where $a=1,2,3$. The constant embedding tensor 
$\Theta_{PQ\vert R}{}^{S}$  describes the embedding of (\ref{eq:gauge}) into $\text{SO}(4,5)$.
Its explicit form is most conveniently described in the basis
\begin{equation}
    X^{P}		\longrightarrow	 \left\{X^{m},\; X_{m},\; X^0,\; X_0,\;X^9\right\}
    \,,\qquad m\in\{1, 2, 3\}\,,
    \label{basis}
\end{equation}
in terms of which the $\text{SO}(4,5)$ invariant tensor takes the form
\begin{equation}	\label{eq:eta}
 		\eta_{PQ}={\small
		\begin{pmatrix}
			0 & \delta_{m}{}^{n} & 0 & 0 &0\\
			\delta^{m}{}_{n} & 0 & 0 & 0 &0\\
			0 & 0 & 0 & 1 & 0\\
			0 & 0 & 1 & 0 & 0\\
			0 & 0 & 0 & 0 & -1
		\end{pmatrix}\,.}
\end{equation}
In this basis, the embedding tensor has the form
\begin{equation}
\label{embedding}
\Theta_{PQ\vert RS}=\Theta_{PQ\vert R}{}^{T}\eta_{TS}=
\theta_{PQRS}+\frac12\Big(\eta_{P[R}\theta_{S]Q}-\eta_{Q[R}\theta_{S]P}\Big)\,,
\end{equation}
in terms of a fully antisymmetric $\theta_{PQRS}$ and a symmetric $\theta_{PQ}$, whose non-vanishing components in the basis (\ref{basis}) are given by \cite{Eloy:2021fhc}
\begin{equation}
	\theta_{0mnp} = 
	\theta_{0m}{}^{np} = 
	\theta_{0m}{}_{n}{}^{p} = 
	\theta_{0mn}{}_{p} = \frac1{2}\,g'\,\varepsilon_{mnp}\,,
 \qquad	\theta_{00}=-4g'\,,
 \label{eq:theta}
\end{equation}
with the three-dimensional coupling constant $g'$ of the first four factors of (\ref{eq:gauge}).
The embedding tensor in the covariant derivatives (\ref{covD}) has an additional component $\theta_{a|MN}$ with non-vanishing components
\begin{equation} 
\theta_{a|MN}\,:\quad \theta_{1|09}=-\theta_{1|90}=4\,g\,,
 \label{eq:theta1}
\end{equation}
in the basis (\ref{basis}). This encodes the couplings induced from the $R$-symmetry gauging in six dimensions. 
The combined embedding tensor (\ref{eq:theta}), (\ref{eq:theta1}), provides a solution to the general quadratic consistency constraint of gauged ${\cal N}=4$ supergravity in three dimensions \cite{deWit:2003ja}. According to the general structure of these theories, covariant derivatives are of the form
\begin{equation}
    D_\mu = \partial_\mu +A_\mu{}^{PQ}\,\Theta_{PQ|RS}\,\mathbb{T}^{RS} + 
    A_\mu^a\,\theta_{a|RS}\,\mathbb{T}^{RS} +A_\mu{}^{RS}\,\theta_{a|RS}\,\mathbb{T}^{a}\,,
    \label{eq:Dcov}
\end{equation}
where the $\mathbb{T}^{PQ}$ denote the generators of ${\rm SO}(4,5)$, and $\mathbb{T}_a$ denote the generators of the ${\rm SO}(3)$ factor of the three-dimensional $R$-symmetry group which exclusively acts in the fermionic sector.

The Chern-Simons term in (\ref{3DL}) is explicitly given by
\begin{align}
{\cal L}_{\rm CS}=\,& -\varepsilon^{\,\mu\nu\rho}\,\Theta_{RS|PQ}\,A_{\mu}{}^{RS}\left(\partial_{\nu}\,A_{\rho}{}^{PQ}  + \frac{1}{3}\, \Theta_{TU|VW}\,f^{PQ,TU}{}_{XY}\, A_{\nu}{}^{VW} A_{\rho}{}^{XY} \right)
\nonumber\\
&
-2\,\varepsilon^{\,\mu\nu\rho}\,\Theta_{a|PQ}\,A_{\mu}{}^{a} \left(\partial_{\nu}\,A_{\rho}{}^{PQ} 
+ \frac{1}{3}\, \Theta_{TU|VW}\,f^{PQ,TU}{}_{XY}\, A_{\nu}{}^{VW} A_{\rho}{}^{XY} 
\right)
,
\end{align}
in terms of the embedding tensor (\ref{eq:theta}), (\ref{eq:theta1}), and the structure constants of $\mathfrak{so}(4,5)$ given by $f^{PQ,RS}{}_{UV} = 4\,\delta_{[U}{}^{[P}\eta^{Q][R}\delta_{V]}{}^{S]}$.

Before describing the scalar potential $V$ in (\ref{3DL}), it is useful to first characterize the scalar field content of the model. The coset (\ref{eq:coset}) is best described in a basis 
\begin{equation}
    X^{P}		\longrightarrow	 \{X^{r},\; X^{I} \}
    \,,\qquad r=1, \dots, 4\,,\;\;\;I = 1, \dots 5\,,
    \label{basisDiag}
\end{equation}
in terms of which the $\text{SO}(4,5)$ invariant tensor takes the diagonal form
\begin{equation}	\label{eq:etaDiag}
 		\eta_{PQ}={\small
		\begin{pmatrix}
			\mathbb{I}_4 & 0 \\
            0 & -\mathbb{I}_5
		\end{pmatrix}\,.}
\end{equation}
The 20 scalar fields of (\ref{3DL}) are associated to the non-compact generators of $\text{SO}(4,5)$, i.e.\ labeled as $\phi^{rI}$, transforming as a bivector under ${\rm SO}(4)_{\rm o}\times {\rm SO}(5)$.
In order to identify their representation content w.r.t.\ the gauge group (\ref{eq:gauge}), we note that the embedding tensor (\ref{eq:theta}) defines the embedding
\begin{equation}
    {\rm SO}(4)_{\rm gauge} = {\rm SO}(3)_{+} \times {\rm SO}(3)_{-} 
    \subset {\rm SO(4)} \times {\rm SO(5)}
    \,,
    \label{gaugeEmb}
\end{equation}
of the gauge group into $\text{SO}(4,5)$, such that the scalar fields fall into representations
\begin{equation}
 \phi^{rI}\,:\;\;   (1,1) \oplus  (1,1) \oplus (3,1)\oplus (3,1)\oplus (1,3)\oplus (3,3)\,, 
\end{equation}
under (\ref{gaugeEmb}). Out of these scalars, 
$(1,1) \oplus (3,1)\oplus (1,3)$ correspond to Goldstone modes associated to the nilpotent generators of the CS gauge group (\ref{eq:gauge}).
The scalar potential $V$ is a function of the remaining 13 scalars, which we write in the form
\be
V= 2\,g'^2 \left( e^{-4\tv} - e^{-2\tv} f(\Phi)\right)+8 g^2 e^{-2\tv}\ .
\label{eq:potential}
\ee
Here $\tv$ denotes the singlet scalar $(1,1)$, and $\Phi$ collectively denotes the remaining 12 scalars $(3,1)\oplus(3,3)$. The explicit form of $f(\Phi)$ was obtained in \cite{Eloy:2021fhc}, and we give it in appendix~\ref{app:3D}.
The contribution in $g^2$ arises directly form the cosmological term in (\ref{L6D}) \cite{Cvetic:2000dm}.

AdS$_3$ vacua correspond to stationary points of the potential (\ref{eq:potential}). 
The structure of (\ref{eq:potential}) shows that the stationary points are defined by the conditions
\begin{equation}
    \partial_\Phi f\big|_{\Phi=\Phi_0} = 0 \,,\qquad 
     e^{-2\tv} =
     \frac12\,f\big|_{\Phi=\Phi_0}-\frac{2\,g^2}{g'^2} \,,
     \label{eq:extremization}
\end{equation}
such that the potential at the extremum takes the value
\begin{equation}
V\big|_{\Phi=\Phi_0}
= 
-2\,g'^2\,e^{-4\tv}
= - \frac2{\ell^2}
\,,
\label{eq:Vl3}
\end{equation}
where $\ell$ denotes the radius of the associated AdS$_3$ metric.
Equations (\ref{eq:extremization}) show that every vacuum $\Phi_0$ of the potential for $g=0$ induces a vacuum of the general potential with arbitrary $g\not=0$, with a shifted value of $\tv$, and the cosmological constant modified according to (\ref{eq:Vl3}). We will use this observation to generalize the one-parameter solution found in \cite{Eloy:2021fhc} to a solution of the general potential with $g\not=0$.
Specifically, this solution
corresponds to switching on a scalar field from the $(3,1)$, thereby breaking the gauge symmetry down to 
\begin{equation}
{\rm U}(2) = {\rm U}(1) \times {\rm SO}(3)_{-} \subset {\rm SO}(3)_{+} \times {\rm SO}(3)_{-}\,.
\label{sym0}
\end{equation}
We review its explicit location $\Phi_0$ in appendix~\ref{app:3D}.
From (\ref{eq:extremization}) we then find a family of solutions for non-vanishing $g$ with only the value of $\tv$ modified to
\begin{equation}
    e^{-2\tv} =
     1-\frac{2\,g^2}{g'^2}\,,
     \label{eq:modDilaton}
\end{equation}
since $f\big|_{\Phi=\Phi_0}=2$\,.
We will now show that the uplift of this solution to six dimensions coincides with the two-parameter solution (\ref{sol})--(\ref{sol3}) above.

To this end, we employ the uplift formulas from \cite{Eloy:2021fhc}, which are not modified by the presence of $g$. In particular, the six-dimensional metric and dilaton are given by
\bea\label{eq: 6dmetric}
	d s^2_6 &=&\Delta^{-2} e^{2\tv} \left( e^{-2\tv} g_{\mu\nu}(x)\, d x^\mu d x^\nu+ h_{mn}\,d y^m d y^n\right)\,,
\\
e^\phi &=&\Delta^{-4}\, e^{4\tilde \varphi}\, ,
\eea
with the three-dimensional metric $g_{\mu\nu}$, and dilaton $\tv$. The symmetric tensor $h_{mn}(y)$ describes the metric on the internal space, and depends exclusively on the three-dimensional scalars $\Phi$ and the $S^3$ spherical harmonics. Its explicit form can be found in \cite{Eloy:2021fhc}. Finally, the warp factor $\Delta$ is given by
\be
\Delta^2= e^{3\tv/2} \left(\frac{{\det h}}{\det \mathring{g}}\right)^{1/4} \,,
\ee
computed from the determinants of the tensor $h_{mn}$ and the metric $\mathring{g}_{m n}$ of the round $S^3$.
The uplift of the above three-dimensional family of AdS$_3$ solutions is then obtained from the results of \cite{Eloy:2021fhc}, upon taking into account the modified value (\ref{eq:modDilaton}) of the three-dimensional dilaton $\tv$, and consequently of the three-dimensional AdS radius (\ref{eq:Vl3}). Putting everything together, we arrive at the following result
\bea
ds^2 &=& e^{-3\omega/2}\,e^{\tv/2} \,g'^{-2} \left(  e^{-2\tv}e^{2\omega}\,g'^2 \ell^2\,ds^2_{{\rm AdS}_3}  +\frac14\,e^{2\omega}\left(\sigma_1^2+\sigma_2^2\right)+\frac14 \sigma_3^2\right)\ ,
\nn\\
F&=& \frac{1}{\sqrt{2}\,g'}\,e^{-\omega}\,\sqrt{e^{2\omega}-1}\, \sigma_1\wedge \sigma_2\ ,
\nn\\
H &=& \frac2{g'^{2}} \,e^{-2\omega}\left( e^{2\tv}e^{2\omega}\,\, \omega_{{\rm AdS}_3} +\omega_{S^3}\right)\ ,\qquad
e^\phi=e^\omega\,e^{\tv}\ ,
\label{ts}
\eea
where the metric $ds^2_{{\rm AdS}_3}$ and volume form $\omega_{{\rm AdS}_3}$ refer to an AdS$_3$ of unit radius. Rescaling the metric and dilaton according to the symmetry (\ref{scaling}) with $\lambda=\mu^{-1}=e^{-3\omega/4}\,e^{\tv/4} \,g'^{-1}$, we precisely recover the solution (\ref{sol})--(\ref{sol3}) upon identification
\begin{equation}
    \xi=e^{2\omega}\,,\quad
    L^2=e^{2\omega}\,e^{2\tv}\,,\quad
    q^2=e^{-4\omega}\,g'^{-4}\,,\quad
    k=\frac{1}{\sqrt{2}\,g'}\,e^{-\omega}\,\sqrt{e^{2\omega}-1}\,,
\end{equation}
where $\tv$ is given by (\ref{eq:modDilaton}).
It is straightforward to check that
this dictionary is compatible with the relations (\ref{qell}) above, as required by the consistency of the truncation. We finally note that the condition for supersymmetry (\ref{susy}) translates into
\begin{equation}
 {\cal N}=(2,0)\;\, {\rm supersymmetry} \;\;\Longrightarrow\;\; \omega=\tv\,,
    \label{susy1}
\end{equation}
in terms of the three-dimensional objects.\footnote{Here, ${\cal N}=(2,0)$ corresponds to 4 real supercharges, and the notation reflects the chiral nature of the vacuum (as opposed to the ${\cal N}=4$ supersymmetry of the Lagrangian).
}

For the following, it will also be useful to introduce the notation
\begin{equation}
\frac{g}{g'}=\frac{\tanh\alpha}{\sqrt2}\;,\qquad
e^\omega = \cosh\beta \,.
\label{eq:alphabeta}
\end{equation}
The parameter $\alpha$ encodes the effect of non-trivial gauging in six dimensions, while $\beta$ keeps track of a non-trivial squashing of the $S^3$.
The condition for supersymmetry (\ref{susy1}) then takes the form 
\begin{equation}
    {\cal N}=(2,0)\;\,{\rm supersymmetry} \;\;\Longrightarrow\;\; \alpha=\beta
    \,.
    \label{eq:susy2}
\end{equation}
A useful relation, that follows from (\ref{eq:modDilaton}) gives the value of the three-dimensional dilaton as
\begin{equation}
e^{2\tv} ={\rm cosh}^2\alpha\,.
\label{tva}
\end{equation}

With this parametrization, and real parameters $\alpha, \beta$, the bound (\ref{eq:bound}) is identically satisfied.
The supersymmetry enhancement point (\ref{susy-enhancement}) with $g=0$ and the background given by  round sphere $S^3$ then corresponds to 
\begin{equation}
    {\cal N}=(4,0)\;\,{\rm supersymmetry} \;\;\Longrightarrow\;\; \alpha=0=\beta
    \,.
    \label{eq:susy4}
\end{equation}

\section{Spectrum at ${\cal N}=(4,0)$ enhancement point and symmetry breaking}
\label{sec:N4}

\subsection{Symmetries and representations}

In order to discuss the full Kaluza-Klein spectrum around this background, it is useful to first 
identify the symmetries that govern and organize the spectrum. 
To begin with, we extend the six-dimensional model (\ref{L6D}) by coupling $p$ additional abelian vector multiplets. In the bosonic action, this amounts to replacing 
\begin{equation}
    e^{\frac12\phi}\,*F\wedge F \,\longrightarrow\,
    e^{\frac12\phi}\,*F\wedge F + e^{\frac12\phi}\,*{\rm F}^i\wedge {\rm F}^i\,,\qquad
    i=1, \dots, p\,.
\end{equation}
From the three-dimensional point of view, the ungauged ${\cal N}=4$ supergravity obtained from toroidal reduction of 6D, ${\cal N}=(1,0)$ supergravity with $p+1$ vector multiplets has the bosonic global symmetries
\begin{equation}
{\rm SO}(4,5+p) \otimes {\rm SO}(3)_{RS}
\;,
\label{eq:sym1}
\end{equation}
where the first factor acts as the isometries on the scalar target space, and the second factor acts exclusively on the fermion fields of the theory (as a chiral factor of the ${\rm SO}(4)$ R-symmetry in three dimensions). 
The 3D supergravity extends the scalar sector of the Lagrangian (\ref{3DL}) to a sigma-model with target space
\begin{equation}
\frac{G}{H} = \frac{{\rm SO}(4,5+p)}{{\rm SO}(4)_{\rm o}\times{\rm SO}(5+p)}
\,,
\label{coset45p}
\end{equation}
and combines it 
with a fermionic sector built from 4 gravitini $\psi_\mu{}^A$ and $4\,(5+p)$ spinors $\chi^{\dot{A}I}$. 
Under the compact subgroup of (\ref{eq:sym1}), the fields of 3D supergravity transform  according to
\begin{equation}
\begin{tabular}{c|cccl}
&${\rm SO}(4)_{\rm o}$ &${\rm SO}(5+p)$&${\rm SO}(3)_{RS}$& \\
\hline
scalars $\phi^{rI}$ & $(\tfrac12,\tfrac12)$ & ${5+p}$ & & $r=1, \dots, 4\,,\;\;\;I=1, \dots, 5+p$\\
fermions $\chi^{\dot{A}I}$ & $(0,\tfrac12)$ & ${5+p}$ & $\tfrac12$ &$A=1, \dots, 4$ \\
gravitini $\psi_\mu^A$  & $(\tfrac12,0)$ && $\tfrac12$ &  $\dot{A}=1, \dots, 4$ \\
\hline
\end{tabular}
\label{eq:fields3Dungauged}
\end{equation}
In addition, the vector fields, gauging the symmetry (\ref{eq:gauge}) are embedded into the adjoint representation of ${\rm SO}(4,5+p)$ according to (\ref{eq:Dcov}). The ${\rm SO}(4)_{\rm gauge}$ part of the gauge group corresponds to the isometries of the round sphere $S^3$. Together with the $R$-symmetry gauging in six dimensions, this breaks the symmetry (\ref{eq:sym1}) down to
\begin{equation}
{\rm SO}(4)_{\rm gauge}  \otimes {\rm SO}(2)_{RS} \otimes {\rm SO}(p+1)
\;,
\label{eq:sym2}
\end{equation}
of which ${\rm SO}(4)_{\rm gauge}$ and ${\rm SO}(2)_{RS}\subset {\rm SO}(3)_{RS}$ are realized as local gauge symmetries within (\ref{eq:gauge}),
whereas ${\rm SO}(p)$ is realized as a global symmetry. The two factors of
${\rm SO}(4)_{\rm gauge}\equiv{\rm SO}(3)_{+}\times{\rm SO}(3)_{-}$ are embedded into the compact groups of 
(\ref{eq:fields3Dungauged}) as
\begin{equation}
{\rm SO}(3)_{+} \stackrel{{\rm diag}}{\subset} {\rm SO}(4)_{\rm o}\;,\qquad
{\rm SO}(3)_{-} \stackrel{{\rm diag}}{\subset} {\rm SO}(4) \subset {\rm SO}(5+p)
\,,
\label{eq:SOLR}
\end{equation}
c.f.~(\ref{gaugeEmb}). Finally, the six-dimensional background further breaks (\ref{eq:sym2}) down to
\begin{equation}
{\rm G}_{\rm spectrum} \;=\; 
{\rm SU}(2) \otimes {\rm U}(1)  \otimes {\rm SO}(2)_{RS} \otimes {\rm SO}(p)
    \;,
\label{eq:sym3}
\end{equation}
due to the squashing of the $S^3$ metric and the flux singling out one among the $n$ vector fields. 
Accordingly, these groups are embedded into (\ref{eq:sym2}), (\ref{eq:SOLR}), as
\begin{equation}
{\rm SU}(2)= {\rm SO}(3)_{-} \;,\quad
{\rm U}(1) \subset {\rm SO}(3)_{+} \;,\quad
{\rm SO}(p)\subset {\rm SO}(p+1)
\,,
\label{eq:embedd}
\end{equation}
c.f.\ (\ref{sym0}).
The full Kaluza-Klein spectrum thus falls into representations of the 
remaining symmetry group ${\rm G}_{\rm spectrum}$, and the AdS$_3$ isometry group ${\rm SO}(2,2)={\rm SL}(2,\mathbb{R})_{L} \times {\rm SL}(2,\mathbb{R})_{R}$.
We denote its representations by
\begin{equation}
    [j]_u^{(q)}(\Delta_L,\Delta_R) 
~=~[j]_u^{(q)} (s\big|\Delta) 
\;,
\label{eq:repsAdS}
\end{equation}
with ${\rm SU}(2)$ spin $j\in\frac12\mathbb{N}$, 
${\rm U}(1)$ charge $u\in\mathbb{Z}$,
${\rm SO}(2)_{RS}$ charge $q=\pm1$,
and 
\begin{equation}
    s\equiv\Delta_R-\Delta_L \;,\qquad \Delta\equiv\Delta_R+\Delta_L
    \;,
\end{equation}
denoting the helicity and the conformal dimensions of the state. The relation between the conformal dimension $\Delta$ and the masses of the various fields around an AdS$_3$ background of radius $\ell$ is given by
\begin{equation}
\Delta\,(\Delta-2) = m^2\ell^2
\;,
\label{eq:mDelta2}
\end{equation}
for fields of spin $s=0$ and $|s|=2$, and
\begin{equation}
\Delta = 1+|m\ell|
\;,
\label{eq:mDelta1}
\end{equation}
for fields of spin $|s|=\frac12$, $|s|=1$, and $|s|=\frac32$. The sign of the helicity $s$ for the latter fields is fixed by that of the mass.

\subsection{${\cal N}=(4,0)$ supersymmetry enhancement point}

Before addressing the higher Kaluza-Klein levels around the background (\ref{sol})--(\ref{sol3}), it is helpful to first consider the supersymmetry enhancement point (\ref{susy-enhancement})
\begin{equation}
\alpha=0=\beta\quad\Longleftrightarrow\quad
g=0=\xi-1\;,
\label{eq:enhancementN4}
\end{equation}
at which the six-dimensional gauging is absent, the internal geometry corresponds to the round $S^3$, and the two-form flux is vanishing. 
Compactifications and spectra around this round geometry have been extensively studied before \cite{deBoer:1998kjm,Deger:1998nm,Samtleben:2019zrh,Eloy:2020uix,Eloy:2021fhc}, and we can use these results to determine the representation content of the Kaluza-Klein spectrum, before computing the associated masses.

Around the AdS$_3\times S^3$ geometry, the internal bosonic symmetries enhance from (\ref{eq:sym3}) to 
\begin{equation}
    {\rm SO}(4)_{\rm gauge} \otimes {\rm SO}(3)_{RS} \otimes {\rm SO}(p)
    ={\rm SO}(3)_+ \otimes{\rm SO}(3)_- \otimes {\rm SO}(3)_{RS} \otimes {\rm SO}(p+1)
    \;,
    \label{eq:symbosN4}
\end{equation}
and supersymmetry enhances to ${\cal N}=(4,0)$. Accordingly, we may label the representations of fields by their spins under (\ref{eq:symbosN4}) and conformal dimensions $(\Delta_L, \Delta_R)$ as
\begin{equation}
[j]_{[k]}^{[z]}(\Delta_L,\Delta_R) 
 \;,
 \label{eq:repsSO4}
\end{equation}
where $j$, $k$, and $z$ now refer to the spin of ${\rm SO}(3)_-$, ${\rm SO}(3)_+$, and ${\rm SO}(3)_{RS}$, respectively. 
Their decomposition into representations (\ref{eq:repsAdS}) follows from the embedding (\ref{eq:embedd}).

The relevant supergroup at the ${\cal N}=(4,0)$ supersymmetry enhancement point is 
\begin{equation}
    {\rm SU}(2|1,1) \otimes {\rm SL}(2,\mathbb{R})_R
    \supset {\rm SO}(3)_+ \times 
    {\rm SL}(2,\mathbb{R})_L \times 
    {\rm SL}(2,\mathbb{R})_R
    \,.
    \label{eq:SU211}
\end{equation} 
This is the small AdS superalgebra in 3D, in which ${\rm SO}(3)_+$ is the $R$-symmetry in 3D not to be confused with the ${\rm SO}(3)_{RS}$ R-symmetry in 6D.
The short multiplets of this superalgebra (in the notation of \cite{Eloy:2020uix}\footnote{W.r.t.\ the conventions of \cite{Eloy:2020uix}, we have to switch chirality $L\leftrightarrow R$ in order to account for an ${\cal N}=(4,0)$ (rather than an ${\cal N}=(0,4)$) vacuum.}) combine the representations
\begin{equation}
{\bf (k+1)}^{(0,j)}_{\Delta_R} \;\;:\;\;  
[j]^{[0]}_{[k/2]}\left(\tfrac{k}2,\Delta_R\right)  \;\oplus\; [j]^{[1/2]}_{[(k-1)/2]}\left(\tfrac{k+1}2,\Delta_R\right)  
\;\oplus\; [j]^{[0]}_{[(k-2)/2]}  \left(\tfrac{k+2}2,\Delta_R\right)
\;,
\label{eq:shortN4}
\end{equation}
with the r.h.s.\ given in the notation of (\ref{eq:repsSO4}), and the notation ${\bf (k+1)}^{(j_1,j_2)}_{\Delta_R}$ where $(j_1,j_2)$ labels $SO(3)_{RS} \times SO(3)_-$, and $(k+1)$ is the dimension of the spin $\frac{k}{2}$ representation of $SO(3)_{+}$.
Since the ${\cal N}=(4,0)$ supersymmetry is chiral, the shortening condition for (\ref{eq:shortN4}) only fixes the left conformal dimensions $\Delta_L$ of the various states as functions of $k$, whereas the right conformal dimensions $\Delta_R$ is unconstrained, but constant throughout the multiplet.

The full Kaluza-Klein spectrum at the ${\cal N}=(4,0)$ supersymmetry enhancement point has been determined in \cite{Eloy:2020uix}. At Kaluza-Klein level $n$, it is given by the following sum of ${\rm SU}(2|1,1) \otimes {\rm SL}(2,\mathbb{R})_R$ short supermultiplets (\ref{eq:shortN4})
\begin{align}
{\rm Spec}_{n} =\,& 
 (\boldsymbol{n+1})^{(0,{n}/{2})}_{{n}/{2}}
  \oplus (\boldsymbol{n+1})^{(0,{(n-2)}/{2})}_{{(n+2)}/{2}}
\oplus(\boldsymbol{n+1})^{(0,{(n+2)}/{2})}_{{(n+2)}/{2}}
\oplus(\boldsymbol{n+1})^{(0,{n}/{2})}_{{(n+4)}/{2}}
   \label{eq:N4spectrum1}
\\[1ex]
 &{}
\oplus(\boldsymbol{n+3})^{(0,{n}/{2})}_{{n}/{2}}
  \oplus(\boldsymbol{n+3})^{(0,{(n-2)}/{2})}_{{(n+2)}/{2}}
\oplus(\boldsymbol{n+3})^{(0,{(n+2)}/{2})}_{{(n+2)}/{2}}
\oplus(\boldsymbol{n+3})^{(0,{n}/{2})}_{{(n+4)}/{2}} 
\;,
\nonumber
\end{align}
built from the fluctuations on the 6D supergravity multiplet, together with $(p+1)$ copies  of the supermultiplets
\begin{equation}
{\rm Spec}^{{\rm vectors}}_{n} =
(\boldsymbol{n+1})^{(0,{n}/{2})}_{{(n+2)}/{2}}\oplus
(\boldsymbol{n+3})^{(0,{n}/{2})}_{{(n+2)}/{2}}
\;,
\label{eq:N4spectrum2}
\end{equation}
describing the fluctuations of the $(p+1)$ six-dimensional vector multiplets. Note that since the two-form flux is vanishing, all of the $p+1$ vector fields coming from six dimensions are on equal footing.
At Kaluza-Klein level $n=0$, the generic formula degenerates, and the spectrum is given by
\begin{equation}
{\rm Spec}_{0} \;\;=\;\; 
{({\bf 3})^{(0,0)}_{0} \;\oplus\;}
({\bf 3})^{(0,1)}_{1} \;\oplus\; ({\bf 3})^{(0,0)}_{2}  \,,\qquad
{\rm Spec}^{{\rm vectors}}_{0} =
({\bf 3})^{(0,0)}_{1} 
\;,
\label{eq:KKlev0SM}
\end{equation}
where the first multiplet is the non-propagating chiral ${\cal N}=(4,0)$ supergravity multiplet.

The six-dimensional vacuum solution (\ref{sol})--(\ref{sol3}) breaks the symmetry down to (\ref{eq:sym3}). Accordingly, the spectrum (\ref{eq:N4spectrum1}), (\ref{eq:N4spectrum2}) breaks into representations of this smaller group.
For $g=0$, the spectrum around the squashed sphere has been computed in \cite{Eloy:2021fhc}, and in the following, we will extend this result to nonvanishing $g$. Note however, that the results of \cite{Eloy:2021fhc} only apply to the bosonic fields of the theory, since, as discussed above after equation (\ref{h6}), the computation was set in another fermionic extension of the bosonic Lagrangian (\ref{L6D}).
For supersymmetric solutions (\ref{susy1}), (\ref{eq:susy2}) half of the supersymmetry is preserved, and the supergroup (\ref{eq:SU211}) breaks down to
\begin{equation}
    {\rm OSp}(2|2) \otimes {\rm SL}(2,\mathbb{R})_R
    \supset {\rm U}(1) \times 
    {\rm SL}(2,\mathbb{R})_L \times 
    {\rm SL}(2,\mathbb{R})_R
    \,.
    \label{eq:OSp22}
\end{equation}

\section{Kaluza-Klein spectrum}
\label{sec:SpectrumKaluza-Klein}

In this section, we will compute the full Kaluza-Klein spectrum around the background  (\ref{sol})--(\ref{sol3}). As discussed in the introduction,
the states of the Kaluza-Klein spectrum are conveniently labeled as $\Phi^{{\cal A}\Lambda}$, where ${\cal A}$ refers to the full field content of the three-dimensional supergravity, and $\Lambda$ labels the scalar harmonics on the round $S^3$. The general Kaluza-Klein mass formula then takes the schematic form \cite{Malek:2019eaz,Malek:2020yue,Cesaro:2020soq,Duboeuf:2023cth}
\begin{equation}
    {\cal M}_{A\Lambda,B\Sigma} = 
    {\cal M}_{AB}\,\delta_{\Lambda\Sigma}
    +{\cal N}_{AB}{}^{PQ} \,{\cal T}_{PQ,\Lambda\Sigma}
    + \delta_{AB}\,{\cal M}_{\Lambda\Sigma}
    \,,
    \label{MM}
\end{equation}
where in particular ${\cal M}_{AB}$ is the mass matrix of the three-dimensional theory around AdS$_3$.
We will thus first compute the three-dimensional mass matrices ${\cal M}_{AB}$, which carry the mass spectrum at the lowest Kaluza-Klein level.
Next, we will work out the remaining contributions to (\ref{MM}) for the fields of different spin, and derive the Kaluza-Klein spectrum at all levels.

\subsection{The lowest KK level}
\label{sec:lowest_KK_level}

The mass matrices for the various fields of three-dimensional model are read off from the couplings of (\ref{3DL}) and its fermionic extension. They are 
expressed in terms of the $T$-tensor of the theory.
The latter is obtained by dressing the embedding tensor (\ref{embedding})--(\ref{eq:theta1}), with the coset representative ${\cal V}$, related as
\begin{equation}
    M_{PQ} = \big({\cal V} {\cal V}^\top\big)_{PQ}
    \,,
\end{equation}
to the scalar matrix $M_{MN}$ in (\ref{3DL}).
Explicitly, the $T$-tensor is then defined as
\begin{align}
    T_{\underline{PQ}\underline{RS}}
    =\,&
    ({\cal V}^{-1})_{\underline{P}}{}^P
    ({\cal V}^{-1})_{\underline{Q}}{}^Q
    ({\cal V}^{-1})_{\underline{R}}{}^R
    ({\cal V}^{-1})_{\underline{S}}{}^S\,
    \theta_{PQRS}
    \,,
    \nonumber\\[.5ex]
    T_{\underline{PQ}}
    =\,&
    ({\cal V}^{-1})_{\underline{P}}{}^P
    ({\cal V}^{-1})_{\underline{Q}}{}^Q
   \,
    \theta_{PQ}\,,\qquad
 T_{a|\underline{PQ}}
    =
    ({\cal V}^{-1})_{\underline{P}}{}^P
    ({\cal V}^{-1})_{\underline{Q}}{}^Q
   \,
    \theta_{a|PQ}
\,.
\label{TTensor}
\end{align}
The bosonic mass matrices of three-dimensional gauged supergravity can be expressed in terms of components of the embedding tensor decomposed in the basis (\ref{basisDiag}). Explicitly, they are given by \cite{Eloy:2020uix,Eloy:2021fhc}
\begin{equation}
{\cal M}^{(1)}_{Ir\vert Js}=-4\,T_{Ir\vert Js}\,,	
\label{eq:M1}
\end{equation}
for the vector fields, and
\begin{align}
	{\cal M}^{(0)}_{Ir\vert{J}s}=\,&
    \delta_{IJ}\left(-\tfrac{8}{3}\,T_{{KLM}r}T_{{KLM}s}+8\,T_{{KL}rp}T_{{KL}sp}-T_{KK}T_{rs}+8\,T\,T_{rs}+T_{{K}r}T_{{K}s}+T_{pr}T_{ps}\right) \nonumber\\
    & \delta_{rs}\left(-\tfrac{8}{3}\,T_{IKLM}T_{JKLM}+8\,T_{{IKL}p}T_{{JKL}p} - T_{KK}T_{IJ}+\,T_{IK}T_{JK} +\,T_{{I}p}T_{{J}p}\right) \nonumber\\
    & +16\, T_{{IJK}p}T_{{K}rsp}-16\,T_{{IK}sp}T_{{JK}rp} -2\,T_{{I}r}T_{{J}s}+2\,T_{{I}s}T_{{J}r}+2\,T_{IJ}T_{rs}
    \nonumber\\
    & +
    2 \,\delta_{rs} \left( T_{a|IK}T_{a|JK}-T_{a|It}T_{a|Jt} \right)-2\, \delta_{IJ} \, T_{a|Kr}T_{a|Ks}
    +2\,T_{a|IJ}T_{a|rs}
    \,,
    \label{eq:MM0}
\end{align}
for the scalar fields. The last line is the contribution due to the new components $\theta_{a|PQ}$ of the embedding tensor; for the $R$-symmetry gauging (\ref{eq:theta1}), the last term vanishes. In contrast, the spin-1 mass matrix (\ref{eq:M1}) is not modified by non-vanishing $g$.
Let us recall that the scalar fluctuations obey standard second-order Klein-Gordon equations, whereas the vector fields satisfy first-order topologically massive field equations, due to the absence of a Yang-Mills term in the action (\ref{3DL}).

The fermionic masses of three-dimensional gauged supergravity are read off from the Yukawa couplings. We recall, that the fermionic mass matrices spelled out in \cite{Eloy:2021fhc} refer to the ${\cal N}=8$ model in three dimensions, whereas the present model is an ${\cal N}=4$ model with in particular different representation content for the fermion fields, c.f.~(\ref{eq:fields3Dungauged}).
For the ${\cal N}=4$ extension of the bosonic model (\ref{3DL}), the Yukawa terms read 
\begin{equation}
e^{-1}\,{\cal L}_{\rm Yuk} =     
\frac12 \, A_1^{AB}\,\bar\psi{}^A_\mu\,\gamma^{\mu\nu}\,\psi^B_\nu\, +
A_2^{A,I\dot{A}}\,\bar\psi{}^A_\mu\,\gamma^\mu \chi^{I\dot{A}} +
\frac12 \, A_3^{I\dot{A},J\dot{B}}\, \bar\chi^{I\dot{A}}\chi^{J\dot{B}}
\,,
\label{eq:Yukawa}
\end{equation}
in terms of scalar dependent tensors $A_1$, $A_2$, $A_3$. These are explicitly given by
\begin{align}
    A_1^{AB} &= \frac{1}{12}\,T_{rstu}\epsilon^{rstu}\delta^{AB} +\frac{1}{4}\,T_{rr}\delta^{AB} + \frac{1}{\sqrt{2}}\,T_{a|rs}C_{a,[rs](AB)}\,, \nonumber\\
    A_2^{A,I\dot{A}} &= \frac{1}{3}\,T_{Iust}\epsilon^{ustr}C_{A\dot{A}r} - \frac{1}{2}\,T_{Ir}C_{A\dot{A}r} + \frac{1}{\sqrt{2}}\,T_{a|Ir}C_{a,A\dot{A}r}\,, \nonumber\\
    A_3^{I\dot{A},J\dot{B}} &= 2\,T_{IJ}\delta^{\dot{A}\dot{B}} - \frac{1}{4}\,T_{rr}\delta^{IJ}\delta^{\dot{A}\dot{B}} + \frac{1}{12} \,T_{rstu}\epsilon^{rstu} \delta^{IJ}\delta^{\dot{A}\dot{B}} \nn \\
    &- 8\, T_{IJrs}C_{[rs][\dot{A}\dot{B}]} - 2\sqrt{2} \,T_{a|IJ} C_{a,[\dot{A}\dot{B}]} - \frac{1}{\sqrt{2}} \,\delta^{IJ} T_{a|rs} C_{a,[rs](\dot{A}\dot{B})}\,,
    \label{eq:A123}
\end{align}
in terms of the different components of the scalar dependent $T$-tensor (\ref{TTensor}), decomposed in the basis (\ref{basisDiag}).
The constant tensors $C$ in these expressions are invariant  under the group SO(4)$_{\rm o} \times {\rm SO}(3)_{RS}$ from (\ref{coset45p}) under which the various fields transform as (\ref{eq:fields3Dungauged}). 
They are uniquely defined by their representation structure together with the normalization which we choose as
\begin{align}
   \lvert C_{a,[rs](AB)} \rvert^2= 36 \,,\qquad
   \lvert C_{A\dot{A}r} \rvert^2= 16 \,,\qquad
   \lvert C_{a,A\dot{A}r}\rvert^2= 48 \,,
   \nonumber\\
   \lvert C_{[rs][\dot{A}\dot{B}]}\rvert^2= 3 \,,\qquad
   \lvert C_{a,[\dot{A}\dot{B}]}\rvert^2= 3 \,,\qquad
   \lvert C_{a,[rs](\dot{A}\dot{B})}\rvert^2= 36 \,.
\end{align}

The Yukawa couplings (\ref{eq:A123}) have been defined universally for all 3D gauged supergravities \cite{deWit:2003ja}, but they had not yet been spelled out explicitly for a general ${\cal N}=4$ gauged supergravity. 
Here, we have determined their precise form from the representation content of the various objects, together with the following characteristic identities
\begin{align}
    A_1^{AC}A_1^{BC}-\frac{1}{2}\,A_2^{A,I\dot{A}}A_2^{B,I\dot{A}} &= -\frac{1}{2}\,\delta^{AB}\,V\,, \nonumber\\
    A_1^{AB}A_2^{B,I\dot{A}}-A_2^{A,J\dot{C}}A_3^{J\dot{C},I\dot{A}} &= C_{A\dot{A}r}\,W_{rI}\,,
    \label{susyWard}
\end{align}
often referred to as supersymmetric Ward identities, relating the matrices $A_1$, $A_2$, $A_3$, to the scalar potential $V$ from (\ref{eq:potential}) and its covariant variation $W_{rI}$, defined by (\ref{susyWard}).

The fermion mass matrices in three dimensions are then read off from (\ref{eq:Yukawa}) and given by
\begin{equation}
    {\cal M}^{(3/2)}_{AB} = -A_{1\,AB}\,, \qquad
     {\cal M}^{(1/2)}_{I\dot{A},J\dot{B}} = -A_{3\,I\dot{A},J\dot{B}}\,,
     \label{Mass1232}
\end{equation}
in terms of the Yukawa tensors (\ref{eq:A123}), where the $T$-tensors (\ref{TTensor}) are to be evaluated on the background solution (\ref{V}).

\begin{table}[bt]
 \renewcommand{\arraystretch}{1.2}
 \centering
 \small
 \begin{tabular}{c|c|c|c}
 $s$ & $m\ell$ & $\Delta$ & $\#$ \\ \hline\hline
 0 & $\sqrt{8}$ & 4 & 1 \\
 0 & 0 & 2 & $4+p$ \\
 0& $2\,\cosh\alpha\,\sinh\beta$ & $1+\sqrt{1+4 \cosh ^2\!\alpha \, \cosh ^2\!\beta }$ & $6+2p$
 \\ \hline
  1&2& 3 &1 \\
 $-1$&$-2$& 3 &$4+p$
\\
1&$1+\sqrt{1+4\cosh^2\!\alpha\,\sinh^2\!\beta}$& $2+\sqrt{1+4\cosh^2\!\alpha\,\sinh^2\!\beta}$ & $2$ \\
$-1$&$1-\sqrt{1+4\cosh^2\!\alpha\,\sinh^2\!\beta}$ & $\sqrt{1+4\cosh^2\!\alpha\,\sinh^2\!\beta}$ & $2$
 \\ \hline
$-1/2$  & $\frac12+\cosh(\alpha\pm\beta)$ & $\frac32+\cosh(\alpha\pm\beta)$ &  $12+4p$\\
$1/2$  &$-\frac32-\cosh(\alpha\pm\beta)$& $\frac52+\cosh(\alpha\pm\beta)$ &$2+2$ \\  \hline
$-3/2$  & $\frac12-\cosh(\alpha\pm\beta)$ & $\frac12+\cosh(\alpha\pm\beta)$ &  $2+2$\\
\hline
\end{tabular}
  \caption{Mass spectrum of three-dimensional supergravity. $(s,\Delta,\#)$ denote the helicity, conformal dimensions from (\ref{eq:mDelta2}), (\ref{eq:mDelta1}), and degeneracy, respectively.}
  \label{eq:KKL0}
\end{table}

Evaluating all the mass matrices given above, the resulting masses are collected in Table~\ref{eq:KKL0}, as functions of $\alpha$ and $\beta$, and further translated into conformal dimensions according to  (\ref{eq:mDelta2}), (\ref{eq:mDelta1}). Masses are given in units of the inverse $\rm{AdS}_3$ radius $\ell^{-1}$, given by
\begin{equation}
   \ell^2 = \frac{g'^2}{\left(g'^2-2\,g^2\right)^2}
    =\frac{{\rm cosh}^4\alpha}{g'^2} =\frac{\sinh^2\!\alpha \cosh^2\!\alpha}{2g^2}\,,
    \label{adsr}
\end{equation}
as found in (\ref{eq:Vl3}), (\ref{eq:modDilaton}), above.

At the supersymmetric point (\ref{eq:susy2}), $\alpha=\beta\not=0$, two of the gravitinos become AdS massless ($m\ell=\frac12$), and accordingly, two more (de-Higgsed) fermions
appear with $m\ell=\frac32$\,.
At this point, the fields organize into ${\cal N}=(2,0)$ super-multiplets, under the supergroup  (\ref{eq:OSp22})
\begin{equation}
    {\rm OSp}(2|2) \otimes {\rm SL}(2,\mathbb{R})_R
    \,.
    \label{eq:sgroup}
\end{equation}
The full spectrum at the supersymmetric point is presented in Tables  \ref{tab:KKL0}, \ref{tab:KKL0Vect},
where we have also included the non-propagating vector and graviton multiplets for which $|s|=\Delta$.
At $\alpha=\beta=0$, supersymmetry further enhances to ${\cal N}=(4,0)$. Accordingly, at this point the massive spin-3/2 multiplet of Tables  \ref{tab:KKL0} becomes massless and enhances the ${\cal N}=(2,0)$ supergravity multiplet to the full ${\cal N}=(4,0)$ supergravity multiplet of (\ref{eq:KKlev0SM}), leaving two more (de-Higgsed) fermions and scalars.
The full field content at this point organizes into the ${\cal N}=(4,0)$ supermultiplets (\ref{eq:KKlev0SM}).

\begin{table}[t!]
 \renewcommand{\arraystretch}{1.4}
 \centering
 \small
 \begin{tabular}{cccccc}
	&$\Delta_{L}$ & $\Delta_{R}$ & $\Delta$ & $s$ & $[j]_u^{(q)}$  
 \\\hline\hline
	 &	$0$ & $2$ & $2$ & $2$ & $\big[0\big]_{0}$ \\\hline\hline
	 &	$0$ & $1$ & $1$ & $1$ & $ \big[1\big]_{0} $ \\\hline\hline
	 &	$2$ & $0$ & $2$ & $-2$ & $\big[0\big]_{0}$ \\\hline
	 &	$3/2$ & $0$ & $3/2$ & $-3/2$ & $\big[0\big]_{\pm1}^{(\mp)}$ \\\hline
	 &	$1$ & $0$ & $1$ & $-1$ & $\big[0\big]_{0}  $ 
\\\specialrule{.4mm}{.8pt}{.8pt}\specialrule{.4mm}{1pt}{1pt}
	 &	$2$ & $1$ & $3$ & $-1$ & $ \big[0\big]_{0}\oplus \big[1\big]_{0} $ \\\hline
	 &	 $3/2$ &  $1$ & $5/2$  &   $-1/2$    & $\big[0\big]_{\pm1}^{(\mp)} \oplus \big[1\big]_{\pm1}^{(\mp)}$ \\\hline
     &  $1$ & $1$  & $2$  &   $0$    & $ \big[0\big]_{0}\oplus \big[1\big]_{0}$ 
\\\hline\hline
  &	$2$ & $2$  & $4$  &   $0$    & $ \big[0\big]_{0}$ \\\hline
	 &	 $3/2$ & $2$ & $7/2$ & $1/2$ & $ \big[0\big]_{\pm1}^{(\pm)} $ \\\hline
	 &	$1$ & $2$ & $3$ & $1$ & $ \big[0\big]_{0} $ 
\\\hline\hline
	 &	$1/2+\gamma$ & $-1+\gamma$ & $-1/2+2\gamma$ & $-3/2$ & $\big[0\big]_{\pm1}^{(\pm)}$ \\\hline
	 &	$\gamma$ & $-1+\gamma$ & $-1+2\gamma$ & $-1$ & $ \big[0\big]_{\pm2} $ 
\\\hline\hline
     &	$1/2+\gamma$ & $\gamma$  & $1/2+2\gamma$  &   $-1/2$    & $ \big[1\big]_{\pm1}^{(\pm)}$  \\\hline
     &	$\gamma$ & $\gamma$  & $2\gamma$  &   $0$    & $  \big[1\big]_{\pm2} $ 
\\\hline\hline	%
	 &	$1/2+\gamma$ & $1+\gamma$  & $3/2+2\gamma$  &   $1/2$    & $\big[0\big]_{\pm 1}^{(\pm)}$ 	 \\\hline
	 &	$\gamma$ & $1+\gamma$ & $1+2\gamma$ & $1$ & $ \big[0\big]_{\pm2} $ \\\hline\hline
  \end{tabular}
  \caption{Spectrum at KK level 0 for the supersymmetric vacua with $\gamma \equiv \cosh^2\!\alpha=\cosh^2\!\beta$. Multiplets are separated by double lines. The first three multiplets correspond to non-propagating degrees of freedom. The first two of these transform only under the ${\rm SL}(2,\mathbb{R})_R$ factor of the group \eq{eq:sgroup}.}
  \label{tab:KKL0}
\end{table}

\begin{table}[t!]
 \renewcommand{\arraystretch}{1.4}
 \centering
 \small
 \begin{tabular}{cccccc}
	&$\Delta_{L}$ & $\Delta_{R}$ & $\Delta$ & $s$ & $[j]_q^{(u)} $ \\\hline\hline
&	$2$ & $1$ & $3$ & $-1$ & $ \big[0\big]_{0} $ \\\hline
&	 $3/2$ &  $1$ & $5/2$  &   $-1/2$    & $\big[0\big]_{\pm1}^{(\mp)}$ \\\hline
&	$1$ & $1$  & $2$  &   $0$    & $ \big[0\big]_{0}$ \\\hline\hline
&	$1/2+\gamma$ & $\gamma$  & $1/2+2\gamma$  &   $-1/2$    & $ \big[0\big]^{(\pm)}_{\pm1}  $  \\\hline
&	$\gamma$ & $\gamma$  & $2\gamma$  &   $0$    & $ \big[0\big]_{\pm2}  $ 
\\\hline\hline
  \end{tabular}
  \caption{Additional states from the $p$ additional vector multiplets in the spectrum at KK level 0 for the supersymmetric vacua with $\gamma \equiv \cosh^2\!\alpha=\cosh^2\!\beta$.} 
  \label{tab:KKL0Vect}
\end{table}

\subsection{Higher KK levels}

The mass spectrum of three-dimensional supergravity computed in the last section yields the masses for the states at the lowest level within the full Kaluza-Klein spectrum. We shall now extend this result to all higher Kaluza-Klein levels. 
According to the general structure of the mass matrices (\ref{MM}), we need to combine the result from the lowest Kaluza-Klein level with the scalar harmonics on the round $S^3$. The latter are given by the tower of symmetric traceless tensor representations the isometry group ${\rm SO}(4)_{\rm gauge}$ as
\begin{equation}
{\cal H}	~=~ \bigoplus_{n=0}^{\infty}\Big(\frac n2,\frac n2\Big)
\;\;\stackrel{{\rm SO}(4)\rightarrow {\rm U}(2)}{\longrightarrow}\;\;
\bigoplus_{n=0}^{\infty} \;\bigoplus_{u\,\in\,{\cal P}_n}  \left[\tfrac{n}{2}\right]_{u}
\;,
	\label{eq:tower_harm}
\end{equation}
where we also give the decomposition of these representations under the ${\rm U}(2)$ isometry group of our background. Furthermore, ${\cal P}_{n}$ denotes the set of integers ranging from $-n$ to $n$ in steps of two. 

The action of the mass matrices (\ref{MM}) on the harmonics is implemented by the representation matrices
$(\tau_{PQ}){}^{\Lambda\Sigma}$ which satisfy the algebra 
\begin{equation}
\left[{\tau}_{PQ},{\tau}_{RS}\right]=-\Theta_{PQ|[R}{}^{T}\, {\tau}_{S]T}+\Theta_{RS|[P}{}^{T}\, {\tau}_{Q]T}\,.
\label{Talgebra}
\end{equation}
Explicitly, they can be realized as matrices
\begin{equation}	\label{eq: curlyTkk1}
	({\tau}\,_{m0}){}^{xy}=\sqrt{2}\,\delta^{[x}_{4}\delta^{y]}_{m}\,,		\qquad
	({\tau}\,^{m}{}_{0}){}^{xy}=-\dfrac{1}{\sqrt{2}}\,\varepsilon^{4{m}xy }\,,
    \qquad x,y=1, \dots,4\,,
\end{equation}
when acting on the ${\rm SO}(4)$ vector representation. Their action on higher harmonics can then be constructed via the matrices
\begin{equation}
	({\tau}_{PQ})_{x_1\dots x_n}{}^{ y_1\dots y_n}=n\,({\tau}_{PQ})_{(\!(x_1}{}^{(\!(y_1}\delta_{x_2}^{{y}_2}\dots\delta_{x_n)\!)}^{ {y}_n)\!)}\,,
    \label{Tnn}
\end{equation}
where $(\!(\dots)\!)$ denotes traceless symmetrization.
Alternatively, it may be convenient to represent the ${\cal T}_{PQ}$ as differential operators 
\begin{equation}
    \tau_{PQ} = {\cal Y}^x\,(\tau_{PQ}){}^{xy}\, \partial_y\,,
    \label{Tdiff}
\end{equation}
acting on polynomials $P({\cal Y}^x)$ of the fundamental harmonics ${\cal Y}^x$ with
\begin{equation}
    {\cal Y}^x{\cal Y}^x=1\,,
\end{equation}
from which the matrix representation (\ref{Tnn}) is immediately deduced.
In the mass formulas, these representation matrices will always appear dressed with the coset representative (\ref{V}), which we denote as
\begin{equation}
{\mathcal{T}}_{\underline{PQ}} = 
({\cal V}^{-1})_{\underline{P}}{}^P
({\cal V}^{-1})_{\underline{Q}}{}^Q\,
\tau_{PQ}
\,.
\label{tautau}
\end{equation}

We now present the analysis for each spin sector separately.

\subsubsection{Spin-2}

In Kaluza-Klein reductions, the
spin-2 fluctuations come in the representations of the scalar harmonics (\ref{eq:tower_harm}). In three dimensions, there appear two copies of this tower related to the two helicities of the spin-2 states
\begin{equation}
2\cdot {\cal H}
=
\bigoplus_{n=0}^{\infty} \;\bigoplus_{u\,\in\,{\cal P}_n}  2\,\left[\tfrac{n}{2}\right]_{u}
\;.
\end{equation}
The spin-2 mass formula takes a compact form in terms of the representation matrices (\ref{Talgebra}) as \cite{Malek:2019eaz,Eloy:2020uix}
\begin{equation}
    {\cal M}^{(2)}_{\Lambda\Sigma} =
    - 2\,
    \left({\cal T}_{\underline{PQ}}{\cal T}_{\underline{PQ}}\right)_{\Lambda\Sigma}
    \,.
    \label{mass-spin2}
\end{equation}
For $g=0$, (i.e.\ $\alpha=0$) this spectrum has been computed in \cite{Eloy:2021fhc}, and the normalized masses have been determined as a function of the Kaluza-Klein level $n$ and the ${\rm U}(1)$ charge $u$ as
\begin{equation}
    m^2\,\ell^2\big|_{g=0} = n(2+n)+u^2 \,\sinh^2\!\beta
    \,.
\end{equation}
The effect of a non-vanishing $g$ enters the mass formula (\ref{mass-spin2}) via the dependence of the dilaton $\tv$ on $g$ according to (\ref{eq:modDilaton}). The structure of the background matrix (\ref{V}) shows that this only affects the components $M^{00}=e^{-2\tv}$ and $M_{00}=e^{2\tv}$. Combining this with the structure of the representation matrices (\ref{eq: curlyTkk1}), we find that the presence of $g$ modifies the spin-2 masses by an overall factor $e^{-2\tv}$. At the same time, the AdS radius $\ell$ changes according to (\ref{eq:Vl3}) by a factor of $e^{2\tv}$. Combining these effects, we read off the normalized spin-2 masses for non-vanishing $g$ to be
\begin{equation}
    m^2\,\ell^2 
    = {\rm cosh}^2\alpha\left[n(n+2)+u^2 \,\sinh^2\!\beta\right]
    \,,
\end{equation}
where we have used (\ref{tva}). The associated conformal dimensions are thus given by
\begin{equation}
\Delta=1+\Gamma^{\rm B}_{(n,u)}\,,
\label{eq:DeltaSpin2}
\end{equation}
with 
\begin{equation}
\Gamma^{\rm B}_{(n,u)}=\sqrt{1+\cosh^2\!\alpha\left[n(2+n)+u^2 \,\sinh^2\!\beta\right]}
\;.
\label{Gammanu}
\end{equation}
At $\alpha=0$, this consistently reduces to the value found in \cite{Eloy:2021fhc}.
For later use, let us note that for special values of $u$, these expressions simplify at the supersymmetric point (\ref{eq:susy2})
\begin{align}
\Gamma^{\rm B}_{(n,n+2)}\big|_{\alpha=\beta}=-1+(n+2)\,\gamma
\;,
\nonumber\\
\Gamma^{\rm B}_{(n,n)}\big|_{\alpha=\beta}=1+n\,\gamma
\;,
\label{eq:GammanuExt}
\end{align}
with $\gamma=\cosh^2\!\alpha=\cosh^2\!\beta$.
Let us also note that 
\begin{equation}
\Gamma^{\rm B}_{(n,u)}\big|_{\alpha=0=\beta}
\;=\; n+1
\,,
\label{eq:gammaN4}
\end{equation}
for all values of $u$, at the full supersymmetry enhancement point (\ref{eq:susy4}) corresponding to the round $S^3$ background.

\subsubsection{Spin 3/2}

According to our general discussion, the spin 3/2 fluctuations can be labeled as $\psi_\mu^{A\Sigma}$, where $A=1, \dots, 4$, is the index of the 3D gravitino. Accordingly, the representation content of the spin-3/2 Kaluza-Klein towers is given by
\bea
\left( \;[0]_{+1}^{(\pm)} \oplus  [0]_{-1}^{(\pm)} \;\right)
\,\otimes\; {\cal H}
&=&
\bigoplus_{n=0}^\infty\Bigg(
 \left[\tfrac{n}{2}\right]_{-n-1}^{(\pm)}
 \oplus \left[\tfrac{n}{2}\right]_{n+1}^{(\pm)} 
 \oplus 
 \bigoplus_{u\,\in\,{\cal P}_{n-1}} 2\cdot \left[\tfrac{n}{2}\right]_{u}^{(\pm)}
 \Bigg)
\,,
\eea 
in the notation (\ref{eq:repsAdS}). We recall that ${\cal P}_{n}$ denotes the set of integers ranging from $-n$ to $n$ in steps of two. 
The mass operator on the towers of spin-3/2 fluctuations is given by
\begin{equation}
{\cal M}^{(3/2)}_{A\Sigma,B\Lambda}=-A_1^{AB}\,\delta_{\Sigma\Lambda}+
   8\,{\rm C}_{[rs][AB]}\,
    {\cal T}_{rs,\Sigma\Lambda} 
    \;.
    \label{M32KK}
\end{equation}
Here, the first term is given by the 3D gravitino mass matrix (\ref{Mass1232}), while the second term carries the dressed representation matrix (\ref{tautau}) and is determined by group theory up to its relative coefficient.
As in (\ref{eq:A123}), the constant tensor ${\rm C}_{[rs][AB]}$ is invariant  under the group SO(4)$_{\rm o} \times {\rm SO}(3)_{RS}$, and uniquely defined by its representation structure and normalization, chosen here as $\lvert C_{[rs][{A}{B}]}\rvert^2= 3$\,.
The coset representative ${\cal V}$ is to be evaluated on the background (\ref{V}).
We stress once more, that the gravitino mass matrix used in the analysis of \cite{Eloy:2021fhc} is not applicable here, since the present model is an ${\cal N}=4$ and not an ${\cal N}=8$ supergravity in three dimensions, with in particular different representation content for all fermionic fields.

The spin-3/2 masses can conveniently be extracted from the following operator identity, 
\bea
\left({\cal M}^{(3/2)}\,\ell-\tfrac12\,{\rm Id}\right)^2 &=&
{\rm Id}+\cosh^2\!\alpha
\left({\cal C}_{\rm SO(4)}
+U^2 \sinh^2\!\beta\right)
\nonumber\\
&&{}
+\cosh\beta\sinh\alpha\Big(\cosh\beta\sinh\alpha+2\,Q\,U \sinh\beta\cosh\alpha \Big)
,
\eea
valid on the entire towers of Kaluza-Klein harmonics. In turn, this identity can be derived from (\ref{M32KK}) by representing the ${\cal T}_{MN}$ as differential operators (\ref{Tdiff}).
Here, ${\cal C}_{\rm SO(4)}$ is the Casimir operator acting on the ${\rm SO(4)}$ harmonics $[\frac{n}{2},\frac{n}{2}]$ with eigenvalues $n(n+2)$,  and $U$ and $Q$ denote the ${\rm U}(1)$ generator and the ${\rm SO}(2)_{RS}$ generator of (\ref{eq:sym3}), respectively. From this identity, together with (\ref{eq:mDelta1}), we read off the 
conformal dimensions for the two helicities $s=\pm3/2$, as $\Delta=1+|m\ell|$
\begin{equation}
s=-\frac32\,:\;\;\Delta =
\frac12+\Gamma^{\rm F}_{(n,u,q)}
\,,
\qquad
s=+\frac32\,:\;\;\Delta =
\frac32+\Gamma^{\rm F}_{(n,u,q)}
\,,
\end{equation}
with
\begin{equation}
\Gamma^{\rm F}_{(n,u,q)}=
\sqrt{\,
\big(\Gamma_{(n,u)}^{\rm B}\big)^2+\cosh\beta\sinh\alpha\Big(\cosh\beta\sinh\alpha+2\,q\,u\,\sinh\beta\cosh\alpha\Big)
}
\,.
\label{eq:GammaF}
\end{equation}
and $\Gamma^{\rm B}_{(n,u)}$ from (\ref{Gammanu}).
Note that at the supersymmetry enhancement point $\alpha=\beta$ this expression simplifies to
\begin{equation}
\Gamma^{\rm F}_{(n,u,q)}\big|_{\alpha=\beta} =\Gamma^{\rm B}_{(n,u+q)}\big|_{\alpha=\beta}
\,,\qquad(q=\pm1)\,,
\label{eq:relationGBF}
\end{equation}
 which will be crucial in identifying the supermultiplet structure below. On the other hand, for $\alpha=0$, we note that
 \begin{equation}
   \Gamma^{\rm F}_{(n,u,q)}\big|_{\alpha=0} =\Gamma^{\rm B}_{(n,u)}\big|_{\alpha=0}
\,,  
 \end{equation}
thus in particular
\begin{equation}
\Gamma^{\rm F}_{(n,u,q)}\big|_{\alpha=0=\beta}
\;=\; n+1
\,,
\label{eq:gammaFN4}
\end{equation}
at the full supersymmetry enhancement point (\ref{eq:susy4}), similar to (\ref{eq:gammaN4}). At this point, the mass spectrum of the spin-3/2 states reproduces the result given in (\ref{eq:N4spectrum1}).

\subsubsection{Spin-1}

The mass matrix for the spin-1 fluctuations has been determined in \cite{Eloy:2020uix} and is not modified by the presence of the additional components of the embedding tensor (\ref{eq:theta1}) for non-vanishing~$g$. It is given in terms of the $T$-tensor (\ref{TTensor}) and the dressed representation matrix (\ref{tautau}) as
\begin{equation}
{\cal M}^{(1)}_{Ir\Lambda\vert Js\Sigma}=
-4\,T_{Ir\vert Js}\,\delta_{\Lambda\Sigma}
-4\,\delta_{rs}{\cal T}_{IJ,\Lambda\Sigma} 
+4\,\delta_{IJ}{\cal T}_{rs,\Lambda\Sigma} 
\,.
\label{eq:M1KK}
\end{equation}
The spin-1 fluctuations of six-dimensional supergravity give rise to the following ${\rm U}(2)$ representations 
\begin{equation}	\label{eq:spectrumVector}
		 \bigoplus_{u\,\in\,{\cal P}_{n}} 2\cdot\Big( \left[\tfrac{n-2}{2}\right]_{u}  \oplus\left[\tfrac{n+2}{2}\right]_{u} \Big)
   \oplus \bigoplus_{u\,\in\,{\cal P}_{n+2}} 2\cdot\left[\tfrac{n}{2}\right]_{u} \oplus \bigoplus_{u\,\in\,{\cal P}_{n-2}} 2\cdot \left[\tfrac{n}{2}\right]_{u} \,,
\end{equation}
at Kaluza-Klein level $n$.  In addition, each of the abelian 6D vector multiplets gives rise to additional states
\begin{equation}	\label{eq:spectrumVectorP}
		 \bigoplus_{u\,\in\,{\cal P}_{n}} 2\cdot\left[\tfrac{n}{2}\right]_{u}  \,,
\end{equation}
Evaluating the mass operator (\ref{eq:M1KK}) on the lowest Kaluza-Klein levels, we observe an intriguing structure for the mass spectrum: the resulting masses precisely take the form found in \cite{Eloy:2021fhc} for $g=0$, simply with all $\Gamma$'s now given by $\Gamma^{\rm B}_{(n,u)}$ of (\ref{Gammanu}). 
This is exactly the same pattern as for the spin-2 fluctuations in (\ref{eq:DeltaSpin2}). Explicitly, this correspond to conformal dimensions
\bea
\Delta=\left\{\Gamma^{\rm B}_{(n,u)},\, 2+\Gamma^{\rm B}_{(n,u)}\right\}
\,.
\eea
In order to identify the correct assignment among these two values for the conformal dimension of a given state, it is most convenient to make use of the limit (\ref{eq:gammaN4}) of $\Gamma^{\rm B}_{(n,u)}$ and match the result against the spectrum (\ref{eq:N4spectrum1}), (\ref{eq:N4spectrum2}) at the ${\cal N}=(4,0)$ enhancement point.

\subsubsection{Spin-1/2}

The mass operator on the tower of spin-1/2 fluctuations is given by
\begin{equation}
{\cal M}^{(1/2)}_{\dot{A}I\Sigma,\dot{B}J\Lambda}=-A_3^{\dot{A}I,\dot{B}J}\,\delta_{\Sigma\Lambda}-8\,\delta^{IJ}\,{\rm C}_{[rs][\dot{A}\dot{B}]}\,
    {\cal T}_{rs,\Sigma\Lambda} 
   -8\,\delta_{\dot{A}\dot{B}}\,
    {\cal T}_{IJ,\Sigma\Omega} 
    \;.
    \label{eq:M12KK}
\end{equation}
The first term is given by the 3D fermion mass matrix (\ref{Mass1232}), while the remaining terms carry the dressed representation matrix (\ref{tautau}) and are determined by group theory up to their relative coefficients. The constant tensor ${\rm C}_{[rs][\dot{A}\dot{B}]}$ has been introduced in (\ref{eq:A123}) above. Again, we recall that for the fermion mass matrices we cannot employ the ${\cal N}=8$ formulas from \cite{Eloy:2021fhc}, but have to directly establish the ${\cal N}=4$ formulas.

The spin-1/2 fluctuations organize into the following ${\rm U}(2)$ representations 
\begin{equation}	
	\begin{aligned}
		& \bigoplus_{u\,\in\,{\cal P}_{n+1}} \Big(\left[\tfrac{n-2}{2}\right]^{(\pm)}_{u} \oplus  \left[\tfrac{n}{2}\right]^{(\pm)}_{u} \oplus\left[\tfrac{n+2}{2}\right]^{(\pm)}_{u} \Big)
   \oplus \bigoplus_{u\,\in\,{\cal P}_{n-1}} \Big( \left[\tfrac{n-2}{2}\right]^{(\pm)}_{u} \oplus  \left[\tfrac{n}{2}\right]^{(\pm)}_{u} \oplus\left[\tfrac{n+2}{2}\right]^{(\pm)}_{u}\Big) \,,
	\end{aligned}
\end{equation}
together with $p$ copies of
\begin{equation}	\label{eq:spectrumFermions}
		 \bigoplus_{u\,\in\,{\cal P}_{n-1}} \left[\tfrac{n}{2}\right]_{u} \quad \text{and} \quad \bigoplus_{u\,\in\,{\cal P}_{n+1}}\left[\tfrac{n}{2}\right]_{u}  \,,
\end{equation}
descending from the vector multiplets in six dimensions.

Evaluating the mass operator (\ref{eq:M12KK}) on the lowest Kaluza-Klein levels, we observe a structure similar to the spin-3/2 masses computed above: the dependence of all spin-1/2 masses on the Kaluza-Klein level $n$ and the ${\rm U}(1)$ charge $u$ is entirely encoded in the $\Gamma^{\rm F}_{(n,u,q)}$ from (\ref{eq:GammaF}). Explicitly, we find conformal dimensions
\bea
\Delta=\left\{-\frac12+\Gamma^{\rm F}_{(n,u,q)},  \frac32+\Gamma^{\rm F}_{(n,u,q)}\right\}
\;,
\nonumber
\eea
for states of helicity $s=-1/2$, and conformal dimensions
\bea
\Delta=\left\{ \frac12+\Gamma^{\rm F}_{(n,u,q)},  \frac52+\Gamma^{\rm F}_{(n,u,q)}\right\}
\;,
\nonumber
\eea
for states of helicity $s=+1/2$, respectively. Again, the most straightforward way of identifying the conformal dimension of a given state follows from comparing its mass at the ${\cal N}=(4,0)$ enhancement point (\ref{eq:N4spectrum1}) to
the limit (\ref{eq:gammaFN4}).

\subsubsection{Spin-0}

Finally, the mass matrix of the spin-0 fluctuations can be read off from \cite{Eloy:2020uix,Eloy:2021fhc}, specifically from equations (4.10)--(4.12) of
\cite{Eloy:2021fhc}, after replacing their equation (4.11), the contribution from 3D supergravity by formula (\ref{eq:MM0}) above, in order to account for the additional $g$-dependent contributions form the scalar potential.

The spin-0 fluctuations organize into the following ${\rm U}(2)$ representations 
\begin{equation}	\label{eq:spectrumSpin0}
	\begin{aligned}
		& \bigoplus_{u\,\in\,{\cal P}_{n}} \Big( \left[\tfrac{n}{2}\right]_{u} \oplus \left[\tfrac{n}{2}\right]_{u}  \Big)
   \oplus \bigoplus_{u\,\in\,{\cal P}_{n+2}} \Big( \left[\tfrac{n-2}{2}\right]_{u} \oplus\left[\tfrac{n+2}{2}\right]_{u} \Big) \nonumber\\
&   \oplus \bigoplus_{u\,\in\,{\cal P}_{n-2}} \Big( \left[\tfrac{n-2}{2}\right]_{u} \oplus\left[\tfrac{n+2}{2}\right]_{u} \Big) \,,
	\end{aligned}
\end{equation}
together with $p$ copies of
\begin{equation}	\label{eq:spectrumSpin0p}
		 \bigoplus_{u\,\in\,{\cal P}_{n-2}} \left[\tfrac{n}{2}\right]_{u} \quad \text{and} \quad \bigoplus_{u\,\in\,{\cal P}_{n+2}}\left[\tfrac{n}{2}\right]_{u}  \,,
\end{equation}
descending from the vector multiplets in six dimensions.

Evaluating the spin-0 mass operator on the lowest Kaluza-Klein levels, we find the same structure as for the fields of higher spin above, the resulting masses precisely take the form found in \cite{Eloy:2021fhc} for $g=0$, simply with all $\Gamma$'s replaced by $\Gamma^{\rm B}_{(n,u)}$ of (\ref{Gammanu}). 
Explicitly, this correspond to conformal dimensions of the form
\begin{equation}
\Delta=\left\{ -1+\Gamma^{\rm B}_{(n,u)}, \, 1+\Gamma^{\rm B}_{(n,u)},\, 3+\Gamma^{\rm B}_{(n,u)}\right\}
.
\end{equation}

\section{Spectrum and multiplet structure}
\label{sec:spectrum}

We have in the previous section determined the full Kaluza-Klein mass spectrum of 6D minimal gauged supergravity around the two-parameter background (\ref{sol})--(\ref{sol3}), by using the consistent truncation to 3D gauged supergravity and the general mass formulas obtained from the ExFT formulation of the model.
In this section, we give a compact summary of the resulting spectrum and of its supermultiplet structure.

Let us start by recalling the two-parameter background solution (\ref{sol})--(\ref{sol3}) of (\ref{L6D}). In terms of the parameters $\alpha$, $\beta$ from (\ref{eq:alphabeta}), in terms of which we have computed the spectrum, this solution takes the form\footnote{This form of the solution can also be obtained directly from 
(\ref{sol})--(\ref{sol3}) by the map: $L=\cosh \alpha \cosh\beta,\hfill\break \xi= \cosh^2 \beta, k=\tanh\alpha \tanh\beta/(2g)$ and $q=\tanh^2\alpha/(g^2 \cosh^2\beta$). The condition \eq{eq:bound} is automatically satisfied for any value of $\alpha$ and $\beta$. 
\label{fn:relation}}
\bea
ds^2 &=&  \left(L^2\, ds^2_{\rm{AdS}_3} +\frac14 \,\sigma_3^2 +\frac14 \cosh^2\!\beta \left(\sigma_1^2+\sigma_2^2\right) \right)\ ,
\label{solab}\\
F&=& \frac1{2g}\,\tanh\alpha\,\tanh\beta\: \sigma_1\wedge \sigma_2\ ,
\nonumber\\
H &=& \frac1{g^2}\,\frac{\tanh^2\!\alpha}{\cosh^2\!\beta} \left( L^2\, \omega_{\rm{AdS}_3} +\omega_{S^3}\right)\ ,\qquad
e^\phi=4\,g^4\,{\coth^4\!\alpha}\,{\cosh^4\!\beta} \ ,
\nonumber
\eea
with $L=\cosh\alpha\,\cosh\beta$\,. Along the line $\alpha=\beta$ in parameter space, this solution, which reduces to that of \cite{Gueven:2003uw},\footnote{The map to the parameters used in the solution of \cite{Gueven:2003uw} is given by $P=b^2=\frac1{8g^2}\frac{\tanh^2\!\alpha}{\cosh^2\!\alpha}$, as well as $k=4ga^2=\frac{1}{2g}\tanh^2\!\alpha$, together with a rescaling (\ref{scaling}) of metric and dilaton with $\lambda=\mu^{-1}=\sqrt{2}\,g\,{\coth\alpha}\,{\cosh\alpha}$.} preserves 4 real supercharges (${\cal N}=(2,0)$ in three dimensions), such that the corresponding spectrum organizes into supermultiplets of the ${\cal N}=(2,0)$ supergroup  (\ref{eq:OSp22}) 
\begin{equation}
    {\rm OSp}(2|2) \otimes {\rm SL}(2,\mathbb{R})_R
    \;.
    \label{eq:OSp22b}
\end{equation}
 At $\alpha=\beta=0$ (which has to be accompanied by sending the gauge coupling $g\rightarrow\frac1{\sqrt2}\,\rm tanh\alpha\rightarrow0$, in order to keep the dilaton in (\ref{solab}) regular), the background reduces to the standard AdS$_3\times S^3$, and supersymmetry is further enhanced. At this point, the full spectrum was given in (\ref{eq:N4spectrum1}), (\ref{eq:N4spectrum2}) in terms of the bigger supergroup ${\rm SU}(2|1,1)$.

In the following, we first briefly review the structure of supermultiplets of the supergroup ${\rm OSp}(2|2)$, before presenting the full Kaluza-Klein spectrum organized in such supermultiplets, and its different limits in parameter space. 

\subsection{${\rm OSp}(2|2)$ supermultiplets}

Representations of the supergroup (\ref{eq:OSp22b}) were studied in \cite{Zhang:2003du}, see also \cite{Gunaydin:1990ag,Lu:2012osp}.
A long ${\cal N}=(2,0)$ multiplet is generated by the action of the two supercharges 
onto a HWS state whose representation we label as $[j]_u (s|\Delta)$ in the notation of (\ref{eq:repsAdS}). Accordingly, we label the full multiplet as $\big\{[j]_u (s|\Delta)\big\}_{\rm L}$. Its full representation content is given by
\begin{gather}    
\label{eq:N2long}
\boxed{
\begin{align}
\big\{[j]_u (s|\Delta)\big\}_{\rm L}\,: \hspace{2.4cm} 
& \qquad\qquad\qquad\quad   [j]_u(s|\Delta)
\nonumber\\[1ex]
& 
[j]_{u+1}^{(-)}(s-\tfrac12|\Delta+\tfrac12) \;\oplus\; [j]_{u-1}^{(+)}(s-\tfrac12|\Delta+\tfrac12) 
\hspace{2.2cm}
\\[1ex]
&\qquad\qquad\quad
[j]_u(s-1|\Delta+1) 
\nonumber
\end{align}
}
\end{gather}
in the notation of (\ref{eq:repsAdS}).
As a consequence of the chiral structure of the supergroup (\ref{eq:OSp22b}), the combination $\Delta_R=\frac12(\Delta+s)$ remains constant throughout the multiplet.
For $u>0$, the multiplet always appears together with its complex conjugate $\big\{[j]_{-u} (s|\Delta)\big\}_{\rm L}$\,, for $u=0$ the multiplet is real.
As usual for long multiplets, in terms of representations of the bosonic subgroup (\ref{eq:sym3}), its field content is
a tensor product of the HWS $[j]_u (s|\Delta)$ with the multiplet $\big\{[0]_{0} (0|0)\big\}_{\rm L}$.

Multiplet shortening for this group happens when the left conformal dimension $\Delta_L=\frac12(\Delta-s)$ saturates the unitarity bound, more precisely, when $\Delta_L$ equals the ${\rm U}(1)$ charge, measured in proper units. As it turns out, in our spectrum the natural unit for the ${\rm U}(1)$ charge is 
$\frac12\gamma = \frac12\cosh^2\!\alpha=\frac12\cosh^2\!\beta$, in terms of the deformation parameters $\alpha$, $\beta$.\footnote{The natural ${\rm U}(1)$ unit follows from evaluating the commutator of supersymmetry transformations.}
Specifically, we find that at this value, the long multiplet (\ref{eq:N2long}) breaks into two short multiplets which we denote as
\begin{equation}
    \big\{[j]_u (s|\Delta)_{\rm L}\big\}\Big|_{\Delta=s+u\,\gamma}
    \quad\longrightarrow\quad
    \big\{[j]_u (s)\big\}_{\rm S}\;\;\oplus\;\;\big\{[j]_{u+1} (s-\tfrac12)\big\}_{\rm S}
    \;.
    \label{eq:shorteningLS}
\end{equation}
The short multiplet $\big\{[j]_u (s)\big\}_{\rm S}$ carries the representation content
\begin{gather}    
\label{eq:N2short}
\boxed{
\begin{align}
\big\{[j]_u (s)\big\}_{\rm S}\,:\hspace{4cm}
&\qquad\;
[j]_{u}(s\big|s+u\,\gamma) 
\nonumber\\[1ex]
&[j]^{(+)}_{u-1}(s-\tfrac12\big|s+u\,\gamma+\tfrac12)
\hspace{4cm}
\end{align}
}
\end{gather}
and for $u>0$ appears together with its complex conjugate $\big\{[j]_{-u}(s)\big\}_{\rm S}$\,.
In \cite{Zhang:2003du} these short multiplets have been referred to as the atypical representations of the supergroup (\ref{eq:OSp22b}). We will see, that the Kaluza-Klein spectrum for the ${\cal N}=(2,0)$ supersymmetric background organizes into infinite towers of both, long and short supermultiplets.

\subsection{Kaluza-Klein spectrum}

We can now present the full Kaluza-Klein spectrum around the two-parameter background (\ref{solab}). A central result of the explicit computation in section \ref{sec:SpectrumKaluza-Klein} is the observation, that the conformal dimensions of all fields are given in compact form in terms of the functions $\Gamma^{\rm B}_{(n,u)}$ from (\ref{Gammanu}) for the bosonic fields, and $\Gamma^{\rm F}_{(n,u,q)}$ from (\ref{eq:GammaF}) for the fermionic fields, respectively. Moreover, in the limit $\alpha=\beta=0$ to the supersymmetry enhancement point, both functions become independent of the ${\rm U}(1)$ charge $u$, and are given by $1+n$ in terms of the Kaluza-Klein level $n$, c.f.\ (\ref{eq:gammaN4}), (\ref{eq:gammaFN4}). Inverting this relation, this offers a direct and efficient way to present the full spectrum. We start from the explicit ${\cal N}=(4,0)$ spectrum (\ref{eq:N4spectrum1}), (\ref{eq:N4spectrum2}), at the supersymmetry enhancement point, break all ${\rm SO}(4)_{\rm gauge} \times {\rm SO}(3)_{RS}$ representations under the bosonic subgroup (\ref{eq:sym3}), and replace in the expressions for the conformal dimensions
\bea
{\rm bosons} &:& n\;\longrightarrow\; -1+\Gamma^{\rm B}_{(n,u)}
\,,
\label{eq:newDeltas}
\\[1ex]
{\rm fermions} &:& n\;\longrightarrow\; -1+\Gamma^{\rm F}_{(n,u,q)}
\,,
\nonumber
\eea
with the functions from (\ref{Gammanu}), (\ref{eq:GammaF}). This yields the entire Kaluza-Klein spectrum around the two-parameter background.

\begin{figure}[h]
\center
\includegraphics[scale=.9]{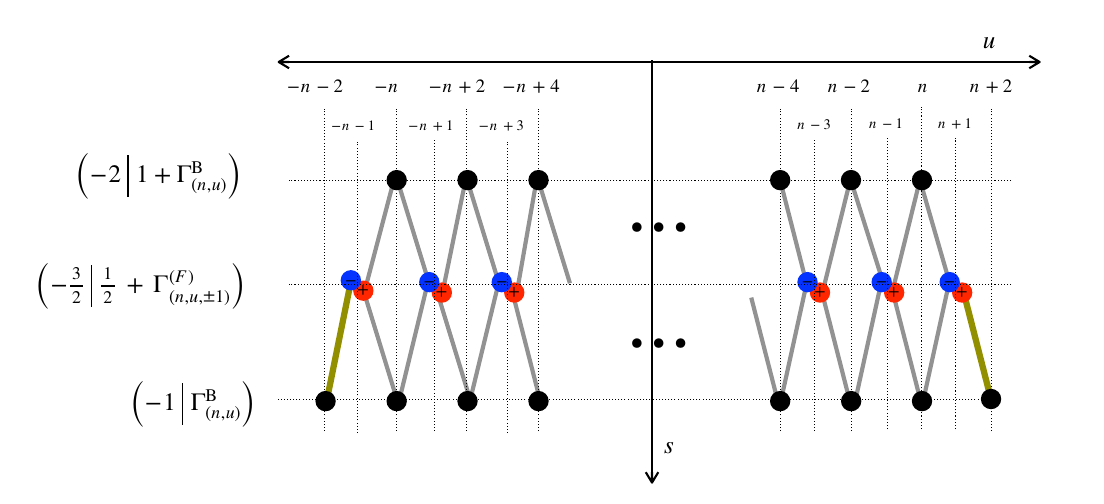}
\caption{
\small States of the ${\cal N}=(4,0)$ spin-2 multiplet $(\boldsymbol{n+3})^{(0,{n}/{2})}_{{n}/{2}}$ decomposed into ${\cal N}=(2,0)$ multiplets. On the vertical axis, we trace the helicity from $-1$ to $-2$, while the horizontal axis indicates the ${\rm U}(1)$ charge $u$, ranging from $-n-2$ to $n+2$. The long multiplets are depicted as diamonds while the green lines at the utmost left and right represent short multiplets.
}
\label{fig:multipletN3}
\end{figure}

As an example, let us illustrate the structure for the states in the ${\cal N}=(4,0)$ spin-2 multiplet $(\boldsymbol{n+3})^{(0,{n}/{2})}_{{n}/{2}}$ in (\ref{eq:N4spectrum1}). Explicitly breaking its field content (\ref{eq:shortN4}) under (\ref{eq:sym3}) yields the following states
\bea
  (\boldsymbol{n+3})^{(0,{n}/{2})}_{{n}/{2}}
&\longrightarrow&
\bigoplus_{u\in{\cal P}_{n}} [\tfrac{n}2]_{u}  \left(-2\big|\,1+\Gamma^{\rm B}_{(n,u)}\right)
\nonumber\\[1ex]
&&{}
\oplus \bigoplus_{u\in{\cal P}_{n+1}} [\tfrac{n}2]^{(\pm)}_{u}
\left(-\tfrac{3}2\big|\,\tfrac{1}2+\Gamma^{\rm F}_{(n,u,\pm1)}\right)  
\nonumber\\[1ex]
&&{}
\oplus   \bigoplus_{u\in{\cal P}_{n+2}} [\tfrac{n}2]_{u}\left(-1\big|\,\Gamma^{\rm B}_{(n,u)}\right)  \;,
\label{eq:breakingN4}
\eea
in the notation of (\ref{eq:repsAdS}), where we have used the prescription (\ref{eq:newDeltas}) in order to fix the conformal dimensions. We recall that ${\cal P}_{n}$ denotes the set of integers ranging from $-n$ to $n$ in steps of two. The structure of states is illustrated in Figure~\ref{fig:multipletN3} on a $(u,s)$ lattice, with black dots representing bosonic states, and red and blue dots representing fermionic states of ${\rm SO}(2)_{RS}$ charge $q=\pm1$, respectively. All states transform in the $[\frac{n}{2}]$ representation under ${\rm SU}(2)$.

At the supersymmetry enhancement point $\alpha=\beta=0$, we have
\begin{equation}
    \Gamma^{\rm B}_{(n,u)}\big|_{\alpha=0=\beta}=\Gamma^{\rm F}_{(n,u,\pm1)}\big|_{\alpha=0=\beta}=1+n\equiv \Gamma_n\,,
\end{equation}
i.e.\ none of the conformal dimensions depend on $u$, and all the states in Figure~\ref{fig:multipletN3} consistently fit into a single multiplet, with conformal dimensions given by $\Gamma_{n}$, $\frac12+\Gamma_n$, and $1+\Gamma_n$, respectively. For non-vanishing $\alpha$ and $\beta$, the conformal dimensions of the various states depend on their ${\rm U}(1)$ charge~$u$ according to (\ref{Gammanu}), (\ref{eq:GammaF}). As a result, the conformal dimensions of bosonic and fermionic fields no longer differ by half integers and can no longer be accommodated in a supermultiplet, consistent with the fact that the solution breaks all supersymmetries for generic values of $\alpha$ and $\beta$. Interestingly, however, we note that even for the non-supersymmetric solution, the masses of the bosonic spin $1$ and the spin $2$ fields are still related, such that their conformal dimensions differ by $1$. We will come back to this observation in the conclusions.

At the ${\cal N}=(2,0)$ supersymmetric line $\alpha=\beta$, the conformal dimensions of neighboring fermionic and bosonic states are related via (\ref{eq:relationGBF})
 \begin{equation}
\Gamma^{\rm F}_{(n,u,\pm1)}\big|_{\alpha=\beta} =\Gamma^{\rm B}_{(n,u\pm1)}\big|_{\alpha=\beta}
\,.
\end{equation}
This relation shows that these states regroup in long ${\cal N}=(2,0)$ supermultiplets (\ref{eq:N2long}) which we depict as diamonds in Figure~\ref{fig:multipletN3}. The conformal dimension of their HWS is given by
\begin{equation}
\Delta = \Gamma_{n,u}\equiv \Gamma^{\rm B}_{(n,u)}\big|_{\alpha=\beta}
\,.
\label{eq:GammaHWS}
\end{equation}
This multiplet structure is a strong consistency test on the masses obtained by explicit computation.
Furthermore, the figure shows that at its left and the right boundary, for $|u|\ge n+1$, there are two left-over states which do not fit into such long multiplets. However, for the bosonic extremal states, relation (\ref{eq:GammanuExt}) implies that
\begin{equation}
    \Delta=-1+(n+2)\,\gamma = s+u\gamma
    \,,
\end{equation}
with $\gamma=\cosh^2\!\alpha=\cosh^2\!\beta$, i.e., these states precisely satisfy the condition for multiplet shortening (\ref{eq:shorteningLS}) and combine into short multiplets (\ref{eq:N2short}) indicated by green lines in Figure~\ref{fig:multipletN3}.

Putting everything together, we thus find the multiplet breaking
\begin{equation}
  (\boldsymbol{n+3})^{(0,{n}/{2})}_{{n}/{2}}
\;\;\longrightarrow\;\;
\big\{ [\tfrac{n}2]_{\pm n}(0)\big\}_{\rm S}  
\;\;\oplus\;\;
 \bigoplus_{u\in{\cal P}_{n}} \big\{ [\tfrac{n}2]_u({-1}|\Gamma_{n,u})\big\}_{\rm L}  
\,,
\end{equation}
of an ${\cal N}=(4,0)$, ${\rm SU}(2|1,1)$ multiplet into a series of long ${\cal N}=(2,0)$ supermultiplets of ${\rm OSp}(2|2)$ together with a short (complex) ${\cal N}=(2,0)$ supermultiplet.
A similar pattern emerges for the decomposition of all the other multiplets in (\ref{eq:N4spectrum1}), (\ref{eq:N4spectrum2}). Summarizing the result, we find that the full Kaluza-Klein spectrum of 6D supergravity around the supersymmetric family of solutions with $\alpha=\beta$, at KK level $n$ is given by the following series of long ${\rm OSp}(2|2)$ multiplets (\ref{eq:N2long})
\bea
{\rm Spec}_{{\rm long}, n}  &=&
\bigoplus_{u\in{\cal P}_{n-2}} \big\{ [\tfrac{n}2]_{u}\left(0|{-1}+\Gamma_{n,u}\right)\big\}_{\rm L}  
\oplus  \bigoplus_{u\in{\cal P}_{n}} \big\{ [\tfrac{n}2]_{u}\left({-1}|\Gamma_{n,u}\right)\big\}_{\rm L}  
\nonumber\\[.5ex]
&&{}
\oplus \bigoplus_{u\in{\cal P}_{n-2}} \big\{[\tfrac{n}2]_{u} \left(2|1+\Gamma_{n,u}\right)\big\}_{\rm L}  
\oplus  \bigoplus_{u\in{\cal P}_{n}} \big\{[\tfrac{n}2]_{u}\left(1|2+\Gamma_{n,u}\right)\big\}_{\rm L}   
\nonumber\\[.5ex]
&&{}
\oplus \bigoplus_{u\in{\cal P}_{n-2}} \big\{[\tfrac{n\pm2}2]_{u}\left(1|\Gamma_{n,u}\right)\big\}_{\rm L}   
\oplus  \bigoplus_{u\in{\cal P}_{n}} \big\{[\tfrac{n\pm2}2]_{u}\left(0|1+\Gamma_{n,u}\right)\big\}_{\rm L}    
\nonumber\\[.5ex]
&&{}
\oplus \;\; 
\bigoplus_{u\in{\cal P}_{n-2}} \big\{[\tfrac{n}2]_{u} \left(1|\Gamma_{n,u}\right)\big\}_{\rm L}  
\oplus  \bigoplus_{u\in{\cal P}_{n}}\big\{[\tfrac{n}2]_{u}  \left(0|1+\Gamma_{n,u}\right)\big\}_{\rm L} 
\;,\qquad
\label{eq:AllSpectrumLong}
\eea
together with the following (complex) short multiplets (\ref{eq:N2short})
\begin{align}
{\rm Spec}_{{\rm short}, n} =\,&
\big\{ [\tfrac{n}2]_{\pm n} (0)\big\}_{\rm S}  
\; \oplus\;  \big\{ [\tfrac{n}2]_{\pm(n+2)}(-1)\big\}_{\rm S}  
\; \oplus\; \big\{[\tfrac{n}2]_{\pm n} (2)\big\}_{\rm S}  
\; \oplus\;  \big\{ [\tfrac{n}2]_{\pm(n+2)}(1)\big\}_{\rm S}  
\label{eq:AllSpectrumShort}
\\[1ex]
&\,{}
 \oplus\, \big\{ [\tfrac{n\pm2}2]_{\pm n} (1)\big\}_{\rm S}   
\, \oplus\,  \big\{ [\tfrac{n\pm2}2]_{\pm (n+2)}(0)\big\}_{\rm S}    
\, \oplus\, 
\big\{ [\tfrac{n}2]_{\pm n} (1)\big\}_{\rm S}  
\, \oplus\,  \big\{ [\tfrac{n}2]_{\pm (n+2)}  (0)\big\}_{\rm S}   
\,.
\nonumber
\end{align}
Here, $\Gamma_{n,u}$ is defined by (\ref{eq:GammaHWS}), and the conformal dimensions of the HWS of the short multiplets are given  by $\Delta=s+u\,\gamma$\,.
We also recall once more that ${\cal P}_{n}$ in these expressions denotes the set of integers ranging from $-n$ to $n$ in steps of two.

In addition, each of the $p$ additional vector multiplets coupled to the 6D supergravity (\ref{L6D}), gives rise to the following supermultiplets
\bea
{\rm Spec}^{\rm vectors}_{{\rm long}, n}  &=&
\bigoplus_{u\in{\cal P}_{n-2}} \big\{[\tfrac{n}2]_{u} \left(1|\Gamma_{n,u}\right)\big\}_{\rm L}  
\oplus  \bigoplus_{u\in{\cal P}_{n}}\big\{[\tfrac{n}2]_{u}  \left(0|1+\Gamma_{n,u}\right)\big\}_{\rm L}
\,,\nonumber\\
{\rm Spec}^{\rm vectors}_{{\rm short}, n}  &=&
\big\{ [\tfrac{n}2]_{\pm n} (1)\big\}_{\rm S}  
\; \oplus\;  \big\{ [\tfrac{n}2]_{\pm (n+2)}  (0)\big\}_{\rm S}   
\,.
\label{eq:spectrumNvectors}
\eea

The structure (\ref{eq:AllSpectrumLong})--(\ref{eq:spectrumNvectors}), of the spectrum is generic for Kaluza-Klein levels $n\ge2$,
whereas at the lowest KK levels $n=0$ and $n=1$ some degeneracies appear.
In particular, for $n=0$ the spectrum is given by the sum of supermultiplets
\bea
{\rm Spec}_{n=0} &=&
  \big\{ [0]_{\pm2}(-1)\big\}_{\rm S}  
\; \oplus\;  \big\{ [0]_{\pm2}(1)\big\}_{\rm S}    
\; \oplus\;  \big\{ [1]_{\pm2}(0)\big\}_{\rm S}     
 \nonumber\\[1ex]
&&{}
\oplus\;  \big\{[0]_{0}(1|3)\big\}_{\rm L}   
\; \oplus\;  \big\{[1]_{0}(0|2)\big\}_{\rm L}    
\;\oplus\;   \big\{[0]_{0}(0|2)\big\}_{\rm L}   
\,,
\label{eq:spectrum0rest}
\\[2ex]
{\rm Spec}^{\rm vectors}_{n=0} &=&
 \big\{ [0]_{\pm2}  (0)\big\}_{\rm S} \;\oplus\; 
  \big\{[0]_{0}(0|2)\big\}_{\rm L}    
  \,,
\label{eq:spectrum0vectors}
\eea
as explicitly presented in Tables~\ref{tab:KKL0} and~\ref{tab:KKL0Vect} above. For $n=1$, it is sufficient to drop from the general expressions (\ref{eq:AllSpectrumLong})--(\ref{eq:spectrumNvectors}) all states of the type $[-\tfrac{1}2]$. We recall that the HWS of a short multiplet comes with conformal dimension $\Delta=s+u\,\gamma$\,.

Let us finally note that, somewhat unexpectedly, at Kaluza-Klein levels $n>0$, the spectrum of fluctuations from the $p$ abelian vector multiplets (\ref{eq:spectrumNvectors}) precisely coincides with the spectrum of fluctuations originating from the distinguished vector multiplet which carries the non-vanishing background according to (\ref{solab}). Explicitly, the spectrum (\ref{eq:spectrumNvectors}) descending from the additional $p$ vector multiplets coincides with the last terms in (\ref{eq:AllSpectrumLong}) and (\ref{eq:AllSpectrumShort}), respectively, which in turn describe the fluctuations around the distinguished background vector multiplet.
Only at the lowest Kaluza-Klein level $n=0$, the spectrum of the $p$ vector multiplets (\ref{eq:spectrum0vectors}) differs from the spectrum of fluctuations around the distinguished vector multiplet contained in (\ref{eq:spectrum0rest}).

\subsection{Limits and scale separation}
\label{sec:limits_scale_separation}

We have computed the full Kaluza-Klein spectrum around the two-parameter solution (\ref{solab}). For the supersymmetric background with $\alpha=\beta$, we have given the result in towers of ${\rm OSp}(2|2)$ supermultiplets in (\ref{eq:AllSpectrumLong})--(\ref{eq:spectrumNvectors}).
In particular, we have shown that all conformal dimensions for bosonic and fermionic fields are linear in the functions
\begin{align}
\Gamma^{\rm B}_{(n,u)}=\,& \sqrt{1+\cosh^2\!\alpha\left[n(2+n)+u^2 \,\sinh^2\!\beta\right]}
\,,\nonumber\\
\Gamma^{\rm F}_{(n,u,q)}=\,&
\sqrt{\,
\big(\Gamma_{(n,u)}^{\rm B}\big)^2+\cosh\beta\sinh\alpha\,\big(\cosh\beta\sinh\alpha+2\,q\,u\,\sinh\beta\cosh\alpha\big)
}
\,,
\label{eq:Gammas}
\end{align}
from (\ref{Gammanu}), (\ref{eq:GammaF}), respectively. As a first non-trivial observation, we conclude that all scalar fields are stable (with masses above the Breitenlohner-Freedman bound), for arbitrary values of $\alpha$ and $\beta$, i.e.\ even for the generic non-supersymmetric background in this family.
Next, we see from (\ref{eq:Gammas}), there is a scale separation in the sense that for $n, u\ne 0$ the ratio of scales
\be
\frac{\ell_{KK}}{\ell} \sim \frac{1}{\cosh\alpha\sinh \beta} \ll 1 \quad {\rm for}\quad \alpha, \beta \gg 1\ ,
\ee
where $\ell_{KK}$ is the length scale scale of the KK excitations. This means that for large $\alpha$ all bosonic masses become very large, except for the states with $n=0=u$\,.  These are the $5+p$ scalar fields and $5+p$ massive vector fields from KK level 0, listed in Table~\ref{eq:KKL0}, together with the 4 massless ${\rm U}(2)$ gauge vectors.
For generic value of $\beta$, all fermionic masses diverge in this limit. The spectrum thus exhibits scale separation, with only finitely many states retaining a non-diverging mass. Moreover, the conformal dimensions corresponding to these remaining fields are all integer as has been observed in other scale separated spectra \cite{Conlon_2022,Apers:2022zjx,Apers_2022,Arboleya:2024vnp,Farakos:2025bwf}. 
 
For the supersymmetric case $\alpha=\beta$, in the same limit of very large  $\alpha$, also the masses of $10+2p$ 
fermionic superpartners from KK level 0 stay finite, as given in Tables~\ref{tab:KKL0}, \ref{tab:KKL0Vect}. The spectrum thus exhibits scale separation even in the supersymmetric case, and the model in this limit reduces to a chiral 3D, ${\cal N}=(2,0)$ theory with $10+2p$ bosonic and fermionic degrees of freedom, coupled to the non-propagating ${\cal N}=(2,0)$ supergravity multiplet. For $p=0$, the finite mass bosons are five topologically massive vectors and five scalars with integer conformal dimensions, and in the supersymmetric case the superpartners have half integer conformal dimensions.

 In the case of $\alpha=\beta$, interestingly it has been observed in \cite{Gueven:2003uw} that in the $\alpha\to \infty$ vacuum solution \eq{solab3-intro} turns into (Minkowski)$_4\times S^2$ solution of \cite{Salam:1984cj}. This can be seen by rescaling (\ref{scaling}) of metric and dilaton with $\lambda=\mu^{-1}=\sqrt{2}\,g\,\cosh\alpha$, redefining $\psi= z\cosh \alpha$, and taking the limit $\alpha\to \infty$. This gives a ${\rm (Minkowski)}_4 \times S^2$ solution \cite{Salam:1984cj} in which the $S^2$ radius is given by $1/(2\sqrt2 g), F=1/(2g), H=0$ and $\phi=1$. For a discussion of the subtleties pertaining to the topological issues, see \cite{Gueven:2003uw}. 

The case $\alpha=0$ corresponds to $g=0$, i.e.\ no gauging in the 6D theory. The background in this case was already found and studied in \cite{Eloy:2021fhc}. Although non-supersymmetric in the present model, the $\alpha=0$ solution is supersymmetric within another supersymmetric extension of the model (to non-chiral 6D, ${\cal N}=(1,1)$ supergravity), as discussed above.
In that case, upon taking the limit $\beta\rightarrow\infty$, the masses of infinitely many fields diverge, however, as seen explicitly from the form of (\ref{eq:Gammas}), the masses of all fields of vanishing ${\rm U}(1)$ charge remain constant along the deformation. We are thus still left with infinitely many fields of masses independent of $\beta$ and comparable to the AdS$_3$ scale, hence there is no scale separation in this case.

At $\beta=0$, the sphere $S^3$ becomes round and consequently the bosonic symmetry enhances to ${\rm SO}(4)$ while supersymmetry is fully broken. Still in the limit $\alpha\rightarrow\infty$, the spectrum exhibits scale separation with $7+p$ scalar fields and $7+p$ massive vector fields from KK level 0 keeping a finite mass, while all other masses diverge.

Finally, at $\beta=0=\alpha$, we recover the known ${\rm AdS}_3\times S^3$ solution of ungauged 6D supergravity, with supersymmetry enhancement to 6D, ${\cal N}=(1,0)$ (equivalently 3D, ${\cal N}=(4,0)$ in three dimensions). The spectrum is given by (\ref{eq:N4spectrum1}), (\ref{eq:N4spectrum2}), with all masses comparable to the AdS$_3$ scale.

\section{Conclusions}
\label{sec:conclusions}

We have in this paper determined the full Kaluza-Klein spectrum of six-dimensional chiral gauged Einstein-Maxwell supergravity around the two-parameter background (\ref{solab}) whose geometry is the direct product of an AdS$_3$ with a squashed three-sphere $S^3$. The computation has employed the consistent truncation of the model to 3D gauged supergravity and the general mass formulas obtained from the ExFT formulation of the model.

We have observed that for all, supersymmetric and nonsupersymmetric backgrounds, all scalar fields are perturbatively stable, i.e.\ exhibit masses above the Breitenlohner-Freedman bound.
Moreover, the spectrum exhibits parametric scale separation: in a particular limit in which $\alpha$ is very large the masses of all but ten bosonic fields become very large. Their conformal dimensions are given by integers, as was observed in other scenarios of scale separation. For the supersymmetric family of backgrounds ($\alpha=\beta$), also ten of the fermionic masses remain finite with half-integer conformal dimensions consistent with the ${\rm OSp}(2|2)$ supermultiplet structure. It will be interesting to revisit and analyze these results within the Swampland Program.

For the supersymmetric family of backgrounds, we have furthermore shown how at generic values of $\alpha=\beta$ the full spectrum organizes into infinite towers of both, long and short supermultiplets of the background isometry supergroup ${\rm OSp}(2|2)$. The understanding of this exact spectrum, particularly the tower of protected short multiplets, should provide further insights for exploring the still elusive holographic CFT dual of the Salam-Sezgin model. We have also noted that in the $\alpha\to \infty $ limit, the vacuum solution \eq{solab3-intro} becomes the (Minkowski)$_4\times S^2$ solution of \cite{Salam:1984cj}, as was found in \cite{Gueven:2003uw}. The KK spectrum of this compactification is known \cite{Salam:1984cj,Aghababaie:2002be}. It would be interesting to study possible implications of this spectrum for the $\alpha\to \infty$ limit of the spectrum we have determined for the vacuum \eq{solab3-intro}. 

Another interesting observation in the spectrum is the following. Even for the generic non-supersymmetric background $\beta\not=\alpha$, the bosonic fields still organize into couples with conformal dimensions that differ by integer values. They could still fit into supermultiplets, however the fermionic fields break this structure. This might be taken as an indication that the solution can actually be embedded into a model with larger supersymmetry, of which some part is preserved, although invisible within the ${\cal N}=(1,0)$ model. Another sign pointing to this is the fact that the solution for $\alpha=0$ breaks all supersymmetries in the ${\cal N}=(1,0)$ model, but is supersymmetric within the ${\cal N}=(1,1)$ model \cite{Eloy:2021fhc}. A candidate theory would be the ${\cal N}=(2,1)$ theory \cite{DAuria:1997caz} into which both, the ${\cal N}=(1,1)$ and the ${\cal N}=(1,0)$ model can be embedded with identical bosonic sector \cite{Roest:2009sn}.


The fact that we have found AdS compactification with parametric scale separation by starting in a dimension lower than 10D, and taking into account $R$-symmetry gauging, is to be contrasted with the result of \cite{Gautason:2015tig} where the Maldacena-Nu\~nez theorem \cite{Maldacena:2000mw} which at the level of 10D supergravity excludes Minkowski and de Sitter flux compactifications (no orientifolds) was generalized to exclude AdS vacua with parametric scale separation. A crucial difference is that in Roman's massive Type IIA supergravity, the sign of the dilaton potential is negative, while it is positive in the Salam-Sezgin model. As noted in \cite{Cvetic:2003xr}, this suggests, in view of \cite{Gibbons:1984kp,Maldacena:2000mw}, that any non-singular internal space must be non-compact. Indeed, the Salam-Sezgin model has been derived from a reduction of pure 10D type I supergravity reduced on a three dimensional non-compact hyperboloid $H(2,2)$, followed by a consistent chiral truncation after a circle reduction \cite{Cvetic:2003xr}. However, this does not provide  a string/M theory origin of the anomaly free extension of this model, since it leaves out Yang-Mills and hypermultiplet couplings needed for anomaly freedom. Nonetheless, it motivates a closer look at the noncompact manifolds in the reductions of 10D and 11D supergravities. One approach for obtaining the highly nontrivial matter field content of the 6D anomaly free $R$-symmetry gauged supergravities, may be to employ the Horava-Witten style bulk+boundary construction by considering the heterotic 10D supergravity on $M^6 \times H_{2,2} \times (S^1/Z_2)$ \cite{Pugh:2010ii}. As noted in \cite{Pugh:2010ii}, reduction on a noncompact space obviously raises important issues which
would need to be addressed to make such a construction physically reasonable. 

Let us alert the reader to the fact that often in the literature the terminology of ``gauged 6D supergravities" is used even if there is no gauging of the $R$-symmetry group. For example, the F theory constructions of gauged  6D supergravities in \cite{Bonetti:2011mw,Grimm:2013fua} lack $R$-symmetry gauging, even though the reader may have an impression to the contrary from \cite{Burgess:2024jkx}. In summary, there is no known F-theory construction of $R$-symmetry gauged 6D supergravity to date. 

Finally, it would be useful to investigate the consequences of our scale separation results in this paper in the context of the AdS swampland conjectures, both in their weak and strong form \cite{Lust:2019zwm}. These are reviewed in \cite{Coudarchet:2023mfs} in considerable detail. Since these are at this stage conjectures, the assumptions and arguments that go into them must, of course, be scrutinized in depth. At the same time, they further motivate the study of the consistency of the anomaly-free $R$-symmetry gauged 6D supergravities beyond the local anomaly freedom, including global and inflow anomalies, and several other swampland conjectures.

\subsection*{Acknowledgments}

We would like to thank Camille Eloy, Gabriel Larios, Yi Pang, and Dimitrios Tsimpis for very helpful discussions.
We thank each other’s home institutions for hospitality during this work.
The work of E.S. is supported in part by NSF grant PHYS-2413006.

\begin{appendix}

\section{The 3D scalar potential and family of stationary points}
\label{app:3D}

The scalar potential of ${\cal N}=4$ gauged supergravity with an embedding tensor (\ref{eq:theta}), (\ref{eq:theta1}), is given by
{\setlength\arraycolsep{1.2pt}
\begin{equation}	\label{eq: scalarpot}
	\begin{aligned}
	 V	&=	\frac1{24}\,\theta_{PQRS}\theta_{TUVW}\Big(M^{TP}M^{UQ}M^{VR}M^{WS}-6\,M^{TP}M^{UQ}\eta^{VR}\eta^{WS}\\
	&\qquad\qquad\qquad\qquad\quad+8\,M^{KP}\eta^{LQ}\eta^{MR}\eta^{NS}-3\,\eta^{KP}\eta^{LQ}\eta^{MR}\eta^{NS}\Big)\\
	&\quad +\frac1{16}\,\theta_{PQ}\theta_{RS}\,\Big(2\,M^{PR}M^{QS}-2\,\eta^{PR}\eta^{QS}-M^{PQ}M^{RS}\Big) \\
    &\quad -\frac12\,\theta_{a,PQ}\theta_{a,RS}M^{PR}\eta^{QS}\,,
	\end{aligned}
\end{equation}
}
The first three lines can be extracted from the general structure of the half-maximal ${\cal N}=8$ result~\cite{Schon:2006kz,Samtleben:2019zrh}. The last term in the potential is the contribution due to the $R$-symmetry gauging in six dimensions. In order to evaluate this potential, 
we choose an explicit parametrization of the ${\rm SO}(5,4)/\big(\text{SO}(4)_{\rm o}\times\text{SO}(5)\big)$ coset representative. After gauge fixing the local $(\mathbb{R}^3\times \mathbb{R}^3)\times \mathbb{R}^1$ shift symmetry of (\ref{eq:gauge}), it may be described in a solvable gauge as
{\setlength\arraycolsep{2pt}
\begin{align}
	\mathcal{V}_{P}{}^{\underline{P}} = 
		\left(
		\begin{array}{c;{2pt/2pt}c;{2pt/2pt}ccc}
			\nu_m{}^n		&	 [(\xi^2+\phi)\,{\nu^\top}]_{{m}{n}}	&	0	&	0		&	-\sqrt2\,\xi_{{m}}		\\\hdashline[2pt/2pt]
			0		& ({\nu^{-1}{}^\top})^{{m}}{}_{{n}}	&	0		&	0	&	0		\\\hdashline[2pt/2pt]
			0	&	0	&	e^{\tv}	&	0	&	0		\\
			0	&	0	&	0		&	e^{-\tv}	&	0\\
			0	&	-\sqrt2\,[\xi^\top\nu^{-1}{}^\top]_{{n}}			&	0	&	0	&	1			\\
		\end{array}
		\right)\,,
\end{align}
}

in terms of the remaining 13 scalar fields. Here, $\phi_{{m}{n}}$ and $\nu_{{m}}{}^{{n}}$ are an antisymmetric and a symmetric $3\times3$ matrix, respectively, with ${\rm det}\,\nu=1$\,. The three-dimensional dilaton is denoted by $\tv$, and $\xi_{{m}}$ is a three-component vector, moreover, $(\xi^2)_{mn}=\xi_m\xi_n$.
With this parametrization, the potential has been computed in~\cite{Eloy:2021fhc} and takes the form (\ref{eq:potential}) with
\begin{align}
 		f(\Phi)  =\,&\frac12\,{\rm Tr}\left(m+m^{-1}\right)-\frac12\,{\rm Tr}\left(\phi m^{-1}\phi\right) +{\rm Tr}\left(\phi m^{-1}\xi^{2}\right)+{\rm Tr}\left(\xi^{2}\right)\\
 		&+\frac12\,{\rm Tr}\left(\xi^{2}m^{-1}\xi^{2}\right)  -\frac{1}{4}\,\det\left(m^{-1}\right)\left(1-{\rm Tr}\left(\phi^{2}\right)-{\rm Tr}\left(\xi^{4}\right)+{\rm Tr}\left(\xi^{2}\right)^{2}\right) \nonumber\\
 		&-\frac{1}{4}\,{\rm T}\left(m^{-1}(\xi^{2}-\phi),(\xi^{2}+\phi)m^{-1},m+(\xi^{2}+\phi)m^{-1}(\xi^{2}-\phi)+2\,\xi^{2}\right)\nonumber\\
 		&-\frac{1}{8}\,{\rm T}\left(m^{-1},m+(\xi^{2}+\phi)m^{-1}(\xi^{2}-\phi)+2\,\xi^{2},m+(\xi^{2}+\phi)m^{-1}(\xi^{2}-\phi)+2\,\xi^{2}\right)\,, 
        \nonumber
 	\end{align}
where ${\rm T}\left(A,B,C\right)=\varepsilon_{mnp}\,\varepsilon_{qrs}\,A^{mq}B^{nr}C^{ps}$, and $m=\nu\nu^\top$\,.
A family of stationary points of this potential for $g=0$ was found in \cite{Eloy:2021fhc} with
\begin{equation}
    \nu={\rm diag}\{1,1,e^{-\omega}\}\,,\qquad \xi=\{0,0,\sqrt{1-e^{-2\omega}}\}\,,\qquad \tv=0=\phi\,.
\end{equation}
According to the discussion after (\ref{eq:Vl3}), this gives rise to a family of solutions for the full potential with nonvanishing $g$, upon shifting the value of $\tv$ according to (\ref{eq:modDilaton}). The coset representative evaluated on the solution is then given by
{\setlength\arraycolsep{2pt}
\begin{align}
	\mathcal{V}_{P}{}^{\underline{P}}=
		\left(
		\begin{array}{ccc;{2pt/2pt}ccc;{2pt/2pt}ccc}
			1	&	0	&	0		&	0	&	0	&	0				&	0			&	0	&0		\\
			0	&	1	&	0		&	0	&	0	&	0				&	0			&	0	&0		\\
			0	&	0	&	e^{-\omega}	&	0	&	0	&	2\,{\rm sinh}\,\omega 		&	0			&0&	-\sqrt2\,\zeta	\\\hdashline[2pt/2pt]
			0	&	0	&	0		&	1	&	0	&	0				&	0			&	0	&0		\\
			0	&	0	&	0		&	0	&	1	&	0				&	0			&	0	&0		\\
			0	&	0	&	0		&	0	&	0	&	e^\omega			&	0			&	0	&0		\\\hdashline[2pt/2pt]
			0	&	0	&	0		&	0	&	0	&	0				&	e^{\tv}	&	0 & 0			\\
		0	&	0	&	0		&	0	&	0	&	0		&		0 &	e^{-\tv}	 & 0			\\
			0	&	0	&	0		&	0	&	0	&	-\sqrt2\,e^\omega\,\zeta	&	0		&	0& 	1			\\
		\end{array}
		\right)\,,
    \label{V}   
\end{align}
}%
with $\zeta=\sqrt{1-e^{-2\omega}}$, and  
$e^{-2\tv} = 1-\frac{2\,g^2}{g'^2}$.

\end{appendix}


\providecommand{\href}[2]{#2}\begingroup\raggedright\endgroup

\end{document}